\documentclass[aps,prb,twocolumn,groupedaddress,longbibliography]{revtex4-1} 
\usepackage{natbib}
\usepackage{graphicx}
\usepackage{epstopdf}
\usepackage{amssymb, amsmath}
\usepackage{multirow}
\usepackage{dcolumn}
\usepackage{bm}
\usepackage{subfigure}
\usepackage{color}
\usepackage{ulem}

\usepackage{xr}
\usepackage[final]{hyperref}
\hypersetup{
        colorlinks=true,       % false: boxed links; true: colored links
        linkcolor=blue,          % color of internal links
        citecolor=blue,        % color of links to bibliography
        filecolor=magenta,      % color of file links
        urlcolor=blue
        }

\newcommand\ME[3]      {\langle{{#1}}|{{#2}}|{{#3}}\rangle} % mtx. element
\newcommand\ket[1]     {|{{#1}}\rangle}
\newcommand\bra[1]     {\langle{{#1}}|}
\newcommand\braket[2]  {\langle{{#1}}|{{#2}}\rangle}

\newcommand\PsiGS      {\Psi_0}

\newcommand\Hop        {{\hat{H}}}
\newcommand\Kop        {{\hat{K}}}

\newcommand\Vop        {{\hat{V}}}

\newcommand\EL         {E_\mathrm{L}^{}}         % local energy functional
\newcommand\Order[1]   {\mathcal{O}\left(#1\right)}

\newcommand\eql[2] % single labeled equation
{
\begin{equation}\label{#1}
\begin{split}
#2
\end{split}
\end{equation}
}

\newcommand\eqsl[1]                            % multiply labeled equation
{
\begin{align}
#1
\end{align}
}

\newcommand\COMMENTED[1] {}

\newcommand\ACCEPT[1] {} % text off
\begin{document}

\title{Local embedding and effective downfolding in the auxiliary-field quantum Monte Carlo method} 
\author{Brandon Eskridge}
\email{bkeskridge@email.wm
.edu}
\affiliation{Department of Physics, William \& Mary, Williamsburg, VA 23187, USA}
\author{Henry Krakauer}
\affiliation{Department of Physics, William \& Mary, Williamsburg, VA 23187, USA}

\author{Shiwei Zhang}
\affiliation{Department of Physics, William \& Mary, Williamsburg, VA 23187, USA}
\affiliation{Center for Computational Quantum Physics, Flatiron Institute, New York, NY 10010, USA}

\begin{abstract}

A local embedding and effective downfolding scheme has been developed and implemented in the auxiliary-field quantum Monte Carlo (AFQMC) method.
A local cluster in which electrons are fully correlated is defined and the frozen orbital method is used on the remainder of the system to construct an effective Hamiltonian which operates within the local cluster.
Local embedding, which involves only the occupied sector, 
has previously been employed in the context of Co/graphene.
Here, the methodology is extended in order to allow for effective downfolding of the virtual sector thus allowing for significant reduction in the computational effort required for AFQMC calculations.
The system size which can be feasibly treated with AFQMC is therefore greatly extended as only a single local cluster is explicitly correlated at the AFQMC level of theory. 
The approximation is controlled by the separate choice of the spatial size of the active occupied region, $R_o$, and of the active virtual region, $R_v$.
The systematic dependence of the AFQMC energy on $R_o$ and $R_v$ is investigated and it is found that relative AFQMC energies of physical and chemical interest
 converge rapidly to the full AFQMC treatment (\textit{i.e.} using no embedding or downfolding). 

\end{abstract}

\maketitle

\section{Introduction}

The solution of the quantum many-electron problem is of fundamental importance to the prediction of the properties of both molecular and condensed matter systems. It is one of the outstanding problems and often a bottleneck in the simulation of materials. 
Due to electron-electron interactions, the computational cost of exact calculations scales exponentially with system size; exact calculations
 are possible for only the smallest systems.
Approximate many-particle methods, such as coupled cluster (CC) methods in quantum chemistry, scale as high-order polynomials in the number of electrons $N$ ( $N^7 - N^8$), which limits their scope. 
Unlike wave-function-based CC and other quantum chemistry approaches, Quantum Monte Carlo (QMC) techniques use stochastic methods to directly compute observables.
The scaling with system size is reduced to $N^3 - N^4$, as in density functional theory or Hartree-Fock methods, making applications to larger systems feasible. Moreover, QMC has demonstrated high accuracy in systems with strong correlated-electron effects.

Many systems of interest include a spatially localized cluster
in which electron correlation effects are expected to be more important than in the rest of the system. 
 It is possible to treat only the local cluster
with a highly accurate, but expensive, computational method while treating the rest of the system with a less expensive, but less accurate, method.
The effective system size at the many-body level of theory is therefore smaller than that of the original system allowing for an overall larger system to be treated.
Local embedding methods were recently reviewed in great detail in the context of a variety of computational methods and embedding strategies~\cite{Gordon2017}.
The auxiliary-field QMC (AFQMC) method~\cite{Zhang2003,AlSaidi2006b,Motta2017},
which scales favorably with system size, as compared with methods capable of a similar degree of accuracy,
 is an ideal tool for embedded calculations.
AFQMC can use any set of one-particle orbitals.
This can be exploited to use unitarily localized orbitals which can be spatially assigned to either the embedding or the host region. 

It was previously shown that local embedding could be employed to reduce the size of the occupied sector in Co/coronene (C$_{24}$H$_{12}$)~\cite{Virgus2014},
leading to highly accurate results directly comparable to experiment.
However, with local embedding alone, the entire virtual space is included at the AFQMC level of theory. 
The virtual space becomes extremely large with high-quality basis sets, especially as the complete basis set limit is approached, which still
limits the size of systems which can be treated using local embedding alone. 
It is therefore desirable to employ an effective local downfolding scheme in order to reduce the size of the virtual sector as well.

In this paper, we show that local embedding, in the occupied space, and local effective downfolding, in the virtual space, can be employed in combination in order to treat strongly correlated local clusters at the AFQMC level of theory at reduced computational cost. Furthermore, the accuracy of embedded and downfolded results can be controlled by the choice of embedding and downfolding 
 parameters. Significant computational savings are achieved in a number of applications while maintaining a high degree of accuracy.

The remainder of the paper is organized as follows. Sec.~\ref{sec:Method} reviews the AFQMC method
and presents a procedure for achieving local embedding combined with local effective downfolding, and concludes with computational details.
A simple application which illustrates the local embedding and downfolding approximation is provided in Sec.~\ref{sec:modelSys}. 
Additionally, the dependence of the systematic errors on the spatial size of the embedding region is studied.
In Sec.~\ref{sec:Application}, local embedding and effective downfolding is applied to Ti-capped carbon chain systems within a graphitic environment.
First, the systematic convergence of the AFQMC energy in the size of the embedding region is studied.
Next, basis-set and finite-size effects in the Ti-capped carbon chain system are considered. 
Finally, it is demonstrated that similar systematics are observed for Ti-capped carbon chain systems which interact with a graphitic environment.
Additional issues and practical limitations of local embedding and downfolding are discussed in Sec.~\ref{sec:Discussion}.
We then conclude with some general remarks in Sec.~\ref{sec:Summary}.

\section{Methodology}
\label{sec:Method}

In this section, we first review key points of the AFQMC method.~\cite{Zhang2003,AlSaidi2006b,Motta2017} We then describe our embedding/downfolding approach, followed by computational details used in the calculations.

\subsection{AFQMC Overview}
\label{sec:GSprojection}

The Hamiltonian in second-quantized form is given by,
\begin{equation}
\label{eq:H0}
    \Hop
  = \Kop + \Vop
  = \sum _{\mu \nu} K_{\mu \nu} c^{\dag}_\mu c_\nu + \frac{1}{2} \sum _{\mu \nu \lambda \rho} V_{\mu \nu \lambda \rho} c^{\dag}_\mu c^{\dag}_\nu c_\lambda c_\rho \, ,
\end{equation} 
where $\Kop$ and $\Vop$ include all one-body and two-body terms, respectively.
The indices run over $M$ orthonormal single-particle basis functions, with
$c^\dag_\mu$ and $c_\mu$ the respective creation and annihilation operators. 
This form of the Hamiltonian applies to 
all approaches for interacting electron systems, from
realistic orbital-based approaches
for materials and molecular systems to lattice-based models.

The ground-state projection AFQMC formalism uses a trial wave function $\ket{\Psi_T}$ from a less-expensive method, such as HF or DFT. 
The resulting canonical orbitals could be used as the orthonormal basis functions in Eq.~(\ref{eq:H0}), 
but for embedding/downfolding, the canonical orbitals are first unitarily transformed to localized basis functions, as described in the next subsection.
Iterative imaginary-time projection is used to obtain the ground state $\ket{\PsiGS}$ from a trial wave function $\ket{\Psi_T}$:
\eql{eq:gs-proj}
{
    \lim_{\beta \to \infty}  e^{-\beta \Hop} \ket{\Psi_T} \approx
    e^{-\tau \Hop}
    e^{-\tau \Hop}
    \cdots
    e^{-\tau \Hop}
    \ket{\Psi_T}
    \to
    \ket{\PsiGS}
    \,,
}
where $\beta$ is the total imaginary projection time, and $\tau$ is a small imaginary time step. 
The projection is cast as a branching random walk with Slater determinants, $\ket{\phi}$.
The ground state energy, $E_0$, is computed using the mixed estimator,
\eql{eq:mixed-Est}
{
 E_0
 =  \frac {\ME{\Psi_T}{\Hop}{\PsiGS} }  {\braket{\Psi_T}{\PsiGS} }
 =  \lim_{\beta \to \infty} \frac {\ME{\Psi_T}{\Hop e^{-\beta \Hop}}{\Psi_T} }  {\ME{\Psi_T}{e^{-\beta \Hop}}{\Psi_T} }
}
The small imaginary time step allows a Trotter-Suzuki decomposition of the imaginary time electron propagator,
\eql{eq:Trotter-Suzuki}
{
    e^{-\tau \Hop}
    \approx
    e^{-\tau \Kop / 2}
    e^{-\tau \Vop}
    e^{-\tau \Kop / 2}
    + \Order{\tau^3}
    \,.
}
The exponential of a one-body operator acting on a Slater determinant simply produces another Slater determinant. $e^{-\tau \Vop}$ can be cast as a high-dimensional intergral over many auxiliary fields, $\bm{\sigma}$, by the application of a Hubbard-Stratonovich transformation \cite{Zhang1997_CPMC,Zhang2003} as given by
\eql{eq:HS1}
{
    e^{-\tau \Hop}
    =
    \int d\bm{\sigma} P(\bm{\sigma})
e^{-\tau {\hat K}/2}
\:e^{\sqrt{\tau} \bm{\sigma} \cdot {\hat{\mathbf v}}}
\:e^{-\tau {\hat K}/2}
    \,,
}
where $P(\bm{\sigma})$ is a normal distribution function and $\hat{\bm{v}}$ is a one-body operator. The operation of the integrand on a Slater determinant $\ket{\phi}$ is given by,
\eql{eq:step}
{
 e^{-\tau {\hat K}/2}
\:e^{\sqrt{\tau} \bm{\sigma} \cdot {\hat{\mathbf v}}}
\:e^{-\tau {\hat K}/2} \ket{\phi}
\equiv e^{-\tau {\hat h}(\bm{\sigma})} \ket{\phi} \to \ket{\phi'}
\,.
}
where $\ket{\phi'}$ is another Slater determinant. An initial population of walkers is prepared (usually set equal to $\ket{\Psi_T}$) and the mixed estimator, given in equation \ref{eq:mixed-Est}, is stochasically sampled. In general, the one-body operator $\hat{h}(\bm{\sigma})$ is complex and as the projection proceeds a complex phase is accumulated for each walker $\ket{\phi}$. This causes statistical fluctuations to increase exponentially with projection time. The phaseless approximation was introduced~\cite{Zhang2003} in order to control this problem. First, an importance function, given by $\braket{\Psi_T}{\phi}$, is introduced and an importance sampling transformation is applied. Eq. \ref{eq:step} then becomes,
\eql{eq:step-imp}
{
e^{-\tau {\hat K}/2}
\:e^{\sqrt{\tau} \left (  \bm{\sigma} - \bm{\bar{\sigma}}[\phi] \right )\cdot {\hat{\mathbf v}}}
\:e^{-\tau {\hat K}/2}  \ket{\phi} \to \ket{\phi'} \, ,
}
 where the force bias, $\bar{\bm{\sigma}}[\phi]$ is given by
\eql{eq:HS-FB}
{
   \bm {\bar{\sigma}}[\phi]
    \equiv
    -\sqrt{\tau}\,
    \frac{\ME{\Psi_T}{\hat{\mathbf v}}{\phi}}{\braket{\Psi_T}{\phi}}
    \,.
}
 The mixed estimator at each time step is then given by a weighted sum over random walkers
\eql{eq:Emixed-walkers}
{
E_0 \approx  \frac {\sum_\phi w_\phi  \EL[\phi]} {\sum_\phi w_\phi} \, ,
}
expressed in terms of the local energy,
\eql{eq:Elocal-def}
{
    \EL[\phi] \equiv \frac{\ME{\Psi_T}{\Hop}{\phi}}{\braket{\Psi_T}{\phi}}
}
of each walker and the weight, $w_\phi$, is accumulated over the random walk. The procedure is described in more detail in Refs.\,[\citenum{Zhang2003,AlSaidi2006b, Purwanto2004,Suewattana2007,Motta2017}].

\subsection{Embedding and Downfolding Approach}
\label{sec:LEAD}

\begin{figure} % Fig. 1
\begin{center} 
\includegraphics[width=0.45\textwidth]{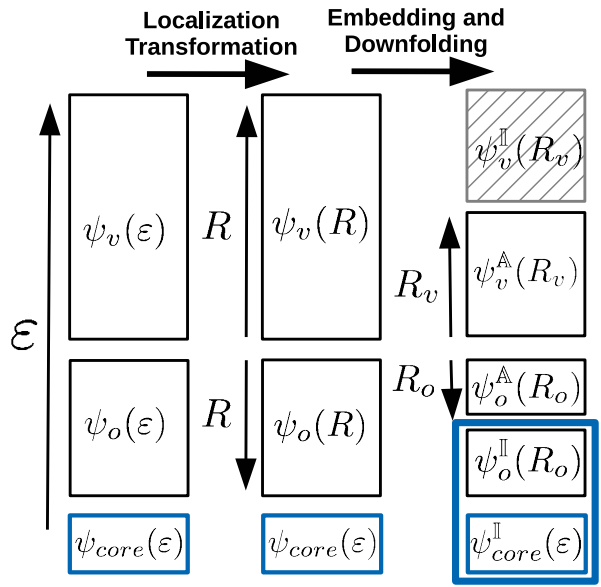}
\caption{\label{fig:EmbeddingSchematic}
Schematic representation of the local embedding and downfolding procedure. 
The leftmost column represents energetically ordered orthonormal core, occupied valence, and unoccupied virtual eigenstates, 
obtained from a lower level of theory such as HF or DFT. 
The occupied $\psi_o(\varepsilon)$ and virtual $\psi_v(\varepsilon)$ orbitals undergo separate unitary transformations to localized orbitals 
$\psi_o(R)$ and $\psi_v(R)$ as indicated in the center column of the figure; $R$ indicates the spatial distance of the orbital from the localized cluster. 
Localized orbitals have been arranged so $R$ increases downward for $\psi_o(R)$ and upward for $\psi_v(R)$.
The rightmost column shows the partitioning of the Hilbert space into active $\mathbb{A}$ and inactive $\mathbb{I}$ sectors (see text).
The cutoffs $R \leq R_o$ and $R \leq R_v$ define, respectively, the occupied and virtual orbitals that are assigned to the active sector $\mathbb{A}$. 
The remaining \mbox{$\psi_o(R \geq R_o)$}  and core orbitals are assigned to $\mathbb{I}$ (blue box); they contribute one-body embedding contributions to the effective
Hamiltonian $\Hop_\mathbb{A}$. The $\psi_v(R \geq R_o)$ are discarded.
}
\end{center}
\end{figure}

Many systems of interest include a spatially localized cluster where electron-correlation effects are expected to be more important than in the rest of the system.
In this section, we describe the partitioning of the Hilbert space into active $\mathbb{A}$ and inactive $\mathbb{I}$ sectors, which 
are used to define an effective Hamiltonian, $\Hop_\mathbb{A}$, to be used in high-accuracy AFQMC calculations.
The construction of  $\Hop_\mathbb{A}$ is schematically represented in Fig.~\ref{fig:EmbeddingSchematic}.
Since $\Hop_\mathbb{A}$ includes only one-body embedded contributions from the $\mathbb{I}$ sector,
 as described below, this greatly extends the system size which can be accurately treated by AFQMC.

The embedding/downfolding effective Hamiltonian, $\Hop_\mathbb{A}$, operates on a reduced Hilbert space, spanned by
the unitarily localized occupied and virtual $\mathbb{A}$ orbitals, 
depicted in the third column of Fig.~\ref{fig:EmbeddingSchematic}. 
$\Hop_\mathbb{A}$ is obtained using a separability approximation for the many-body wave function:~\cite{Kahn1976,HUZINAGA199151}
\begin{equation}
\label{eq:sep}
	\Psi \approx \mathcal{A} (\Psi_\mathbb{I} \Psi_\mathbb{A}) \, ,
\end{equation}
where $\Psi_\mathbb{I}$ and $\Psi_\mathbb{A}$ depend, respectively, only on the $\mathbb{I}$ and $\mathbb{A}$ orbitals in Fig.~\ref{fig:EmbeddingSchematic}.
Both $\Psi_\mathbb{I}$ and $\Psi_\mathbb{A}$ are separately antisymmetric and normalized, and mutually orthogonal;
the antisymmetrizer $\mathcal{A}$ permutes electrons between $\Psi_\mathbb{I}$ and $\Psi_\mathbb{A}$. 
This approximation allows the energy of the total system to be mapped onto an equivalent system which explicitly includes only the $\mathbb{A}$ orbitals: 
\begin{equation}
\label{eq:energyEq}
	E = \bra{\Psi} \Hop \ket{\Psi} = \bra{\Psi_\mathbb{A}} \Hop_\mathbb{A} \ket{\Psi_\mathbb{A}} \, ,
\end{equation}
where $\Hop_{\mathbb{A}}$ is given by \cite{Purwanto2013}:
\begin{equation}
\label{eq:Hdownfold}
	\Hop_{\mathbb{A}} = \sum _{ij \in \mathbb{A}} K_{ij} c^{\dag}_i c_j + \frac{1}{2} \sum_{ijkl \in \mathbb{A}} V_{ijkl} c^{\dag}_i c^{\dag}_j c_k c_l 
	\\ + \sum _{ij\in \mathbb{A}} V^{\mathbb{I} - \mathbb{A}}_{ij} c^{\dag}_i c_j + E^{\mathbb{I}} \, .
\end{equation}
The first two terms in Eq.~\ref{eq:Hdownfold} are one and two-body terms that depend only on the $\mathbb{A}$ orbitals.
The third term is a one-body embedding potential that describes Coulomb and exchange interactions between the $\mathbb{A}$ orbitals
and the environment, which is spanned by the $\mathbb{I}$ orbitals. Formally, the third term is a nonlocal pseudopotential. 
The constant-energy fourth term $E^{\mathbb{I}}$ includes the kinetic and internal interactions of orbitals in $\mathbb{I}$.
The effective Hamiltonian $\Hop_{\mathbb{A}}$ has the same form as Eq.~(\ref{eq:H0}), so AFQMC calculations proceed 
exactly as outlined in Sec.~\ref{sec:GSprojection}. 

A unitary transformation of the occupied and virtual orbitals to localized orbitals is accomplished by
minimizing a cost function that depends on the localized orbitals.
The most used methods are Foster-Boys (FB)~\cite{Boys1960,Boys1966}, Pipek-Mezey (PM)~\cite{Pipek1989} and Edmiston-Ruedenberg (ER)~\cite{Edmiston1963}.
Some methods, such as FB, use the $p^{th}$ moments $\mu^i_p$,  (and its spread $\sigma^i_p$) \cite{Hoyvik2012}:
\begin{equation}
\label{eq:pthMomSpread}
	\mu^i_p = \bra{\phi_i} (\hat{r} - \bra{\phi_i}\hat{r}\ket{\phi_i})^p\ket{\phi_i}; ~~~~ \left( \sigma^i_p = \sqrt[p]{\mu^i_p} \right ),
\end{equation}
with the cost function specified by:
\begin{equation}
\label{eq:costfunction}
	\xi^m_p=\sum_i \left (u^i_p \right )^m.
\end{equation}
The FB method, for example, uses $\xi^1_2$. 
Compared to the occupied manifold, FB, PM and ER have more difficulty optimizing the virtual orbitals. In recent work, the use of
$\xi^1_4$ based on the fourth central moments (FM) has been shown to be effective for localizing virtual orbitals \cite{Hoyvik2012}. 
FM localization can be further improved using $\xi^2_4$ (FM2), with a penalty-exponent $m=2$
 (as previously done for the FB cost function~\cite{Jansik2011}).
FM2 was found to satisfactorily 
achieve a uniformly localized set of virtual orbitals. \cite{Hoyvik2012}. 
FM and FM2 place more weight on the orbital tail and therefore tend to suppress orbitals with long tails.

A more stringent criterion for a set of localized orbitals that has been considered~\cite{Hoyvik2012}  is the spread of the least-local orbital. 
This can be useful in divide-and-conquer local-correlation methods that divide the system into local clusters.\cite{Knizia2013,Yoshikawa2017} 
Here we focus on systems with a single local cluster, so the efficiency of the method is affected only by the localization of virtual orbitals in
the cluster neighborhood and not by those far from the cluster.
	
\subsection{Computational details}
\label{sec:compDets}

In this paper, the canonical occupied and virtual orbitals are obtained from restricted Hartree-Fock (RHF) calculations
using  NWChem~\cite{NWChem-6.0}. The RHF wave functions are also used as
the trial wave function in all AFQMC calculations.
Standard gaussian-type orbital (GTO) cc-pVXZ basis sets\cite{EMSL_BasisSets2007} are used, as described in the sections below. 
AFQMC calculations were done using the generic second-quantized Hamiltonian in Eq.~(\ref{eq:H0}) for the full system and Eq.~(\ref{eq:Hdownfold}) for the active cluster, as described
by Al-Saidi {\it et al.}~\cite{AlSaidi2006b}.

The ERKALE\cite{ERKALE} computer code is used to apply FB localization to the occupied orbitals and FM localization to the virtual orbitals, unless otherwise stated.
Although ERKALE has not implemented, to date, trust-region minimization~\cite{Hoyvik2012-TR,Lehtola2013}, 
which is more robust than gradient-based minimization algorithms (particularly for FM  and FM2 cost functions), 
its localization engine is capable of producing orbitals which are sufficiently local for the purpose of the present work, as indicated in the sections below.

\section{Illustrative Application}
\label{sec:modelSys}

\begin{figure}[h] %Fig. 2
\begin{center}
\includegraphics[width=0.45\textwidth]{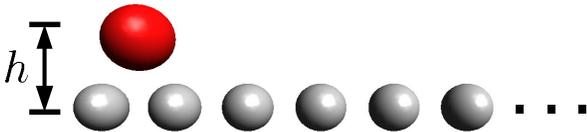}
\caption{\label{O+H20}  O/H$_{20}$ system geometry. The O atom is a distance $h$ above an endpoint H-H bridge site of a 20-atom H chain.
The H-H bond length is fixed at 1.78 Bohr, corresponding approximately to the equilibrium bond length of the H chain.~\cite{Al-Saidi2007}.
The molecular visualization was generated using Avogadro~\cite{Hanwell2012}.
}
\end{center}
\end{figure}

The methodology of the local embedding and effective downfolding method 
is illustrated in this section for a simple model system consisting 
of a single oxygen atom and a linear chain of 20 hydrogen atoms as depicted in Fig.~\ref{O+H20}. The atomic coordinates are given in Table~\ref{sup-tab:O+H20-geom}.
The AFQMC total energy is calculated as a function of $h$, the position of the O atom above the H-H bridge site at one end of the linear chain.
RHF calculations were performed using a cc-pVDZ basis for both O and H atoms. 
The Hamiltonian is represented in the basis of localized RHF orbitals, using FB and FM localization on the occupied and virtual orbitals respectively. 
FM2 localization failed during the line search procedure as implemented in ERKALE~\cite{Lehtola2013} for the virtual orbitals of O/H$_{20}$ with default settings.

\begin{figure} %Fig. 3
\begin{center}
\includegraphics[width=0.45\textwidth]{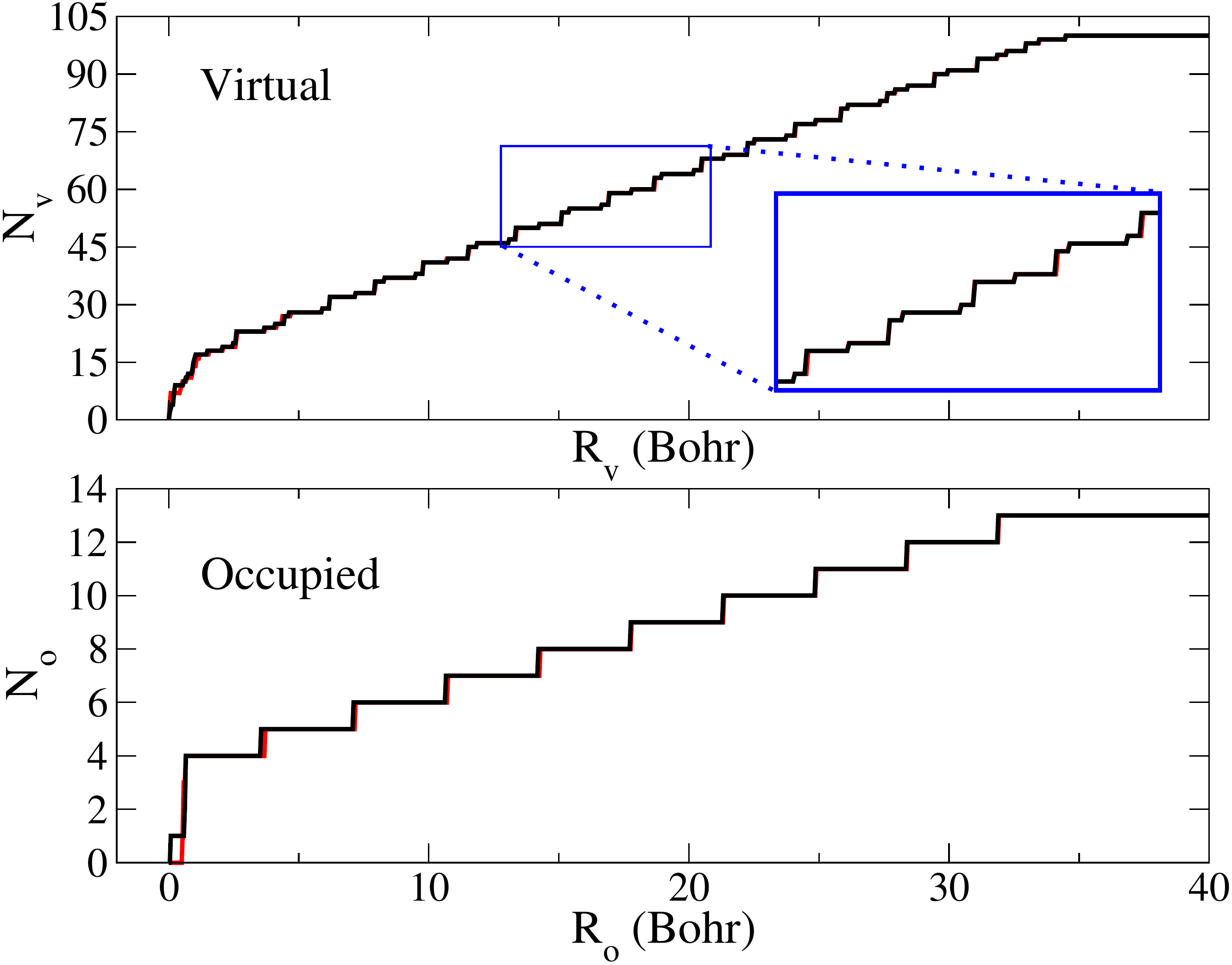}
\caption{\label{fig:O+H20centDist} 
The number of localized virtual (upper panel) and occupied (lower panel) orbitals within the active Hilbert space as a function of the localization 
cutoff radii in H$_{20}$.
In both the upper and lower panels, two data series (nearly indistinguishable) are plotted for O atom heights of 1.81 Bohr, 
corresponding roughly to the equilibrium, 
 (in red) and 3.89 Bohr (in black).
 The step features in the lower panel
 correspond to H-H bond orbitals. The top panel shows smaller step features. However, these features are less physically intuitive.}
\end{center}
\end{figure}

We regard the active region as roughly centered on the O atom and including some neighboring H atoms and define
the localization radii $R_o$ and $R_v$ from the position of the O atom, projected onto the chain, to the centroid of the orbital. 
Fig.~\ref{fig:O+H20centDist} shows the numbers
of occupied and virtual orbitals assigned to $\mathbb{A}$, $N_o$ and $N_v$, as a function 
of $R_o$ and $R_v$, respectively, for two values of $h$. 
The plotted distribution is based on the centroids of the localized orbitals. Beyond about 1-2 Bohr, the distributions for the two
heights are nearly identical. As shown below, this leads to relative insensitivity of total energy differences when $R_o$ and $R_v$ are varied. 
The unit-step feature for the occupied localized orbitals reflects that they are centered on the H-H bond centers away from the O atom.
 
 \begin{figure} %Fig. 4
\begin{center}
\includegraphics[width=0.45\textwidth]{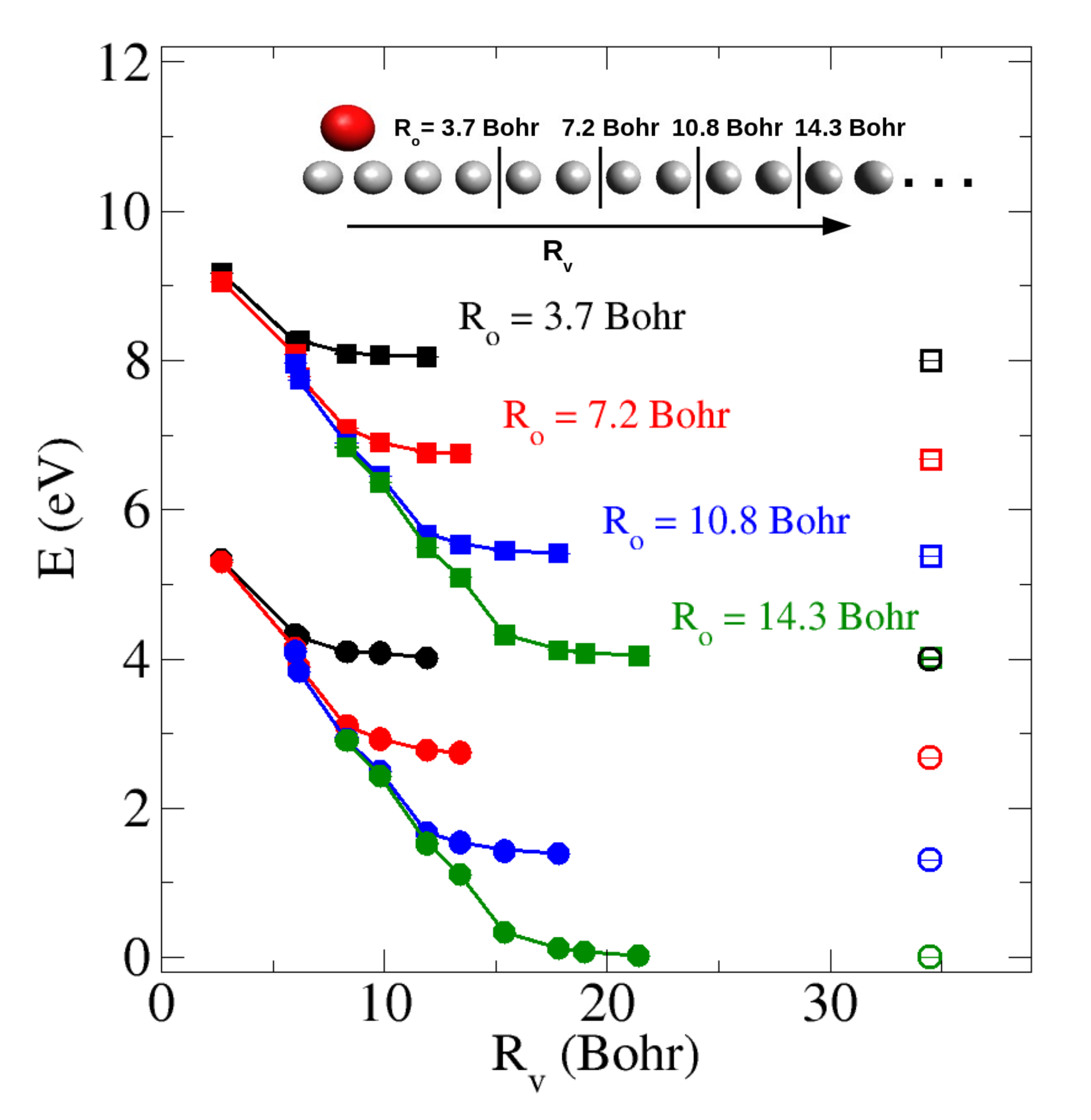}
\caption{\label{O+H20_EvRvirt} O/H$_{20}$ AFQMC \mbox{energy vs.\,$R_{v}$} for four values of $R_o$.  
The inset diagram roughly illustrates the $R_o$ cutoff radii on the system geometry (molecular visualization generated using Avogadro~\cite{Hanwell2012}).
Filled squares are for the O atom at $h=3.89$ Bohr, with $R_o$ values indicated.
Filled circles are for O at the equilibrium height, $h=1.81$ Bohr, using the same $R_o$ color coding.  
Solid lines are a guide to the eye. 
Stochastic uncertainties are smaller than the plot symbols.
Unfilled symbols indicate the full virtual space result (\textit{i.e.} all local virtual orbitals are assigned to $\mathbb{A}$) at $R_v$ corresponding to the full length of the \mbox{H$_{20}$} chain. 
}
\end{center}
\end{figure}

Figure~\ref{O+H20_EvRvirt} displays the calculated AFQMC energy 
of the \mbox{O/H$_{20}$} chain as a function of $R_v$ for four values of $R_o$ and two values of $h$.
All the curves show similar qualitative convergence behavior. 
For \mbox{$R_v < R_o + C$}, where $C \approx 3$ Bohr, the total energy decreases approximately linearly with increasing $R_v$.
For  $R_v > R_o + C$ the total energy is converged very nearly to the full virtual space treatment (\textit{i.e.} all virtual orbitals are included in $\mathbb{A}$) 
for the chosen $R_o$, which is indicated at the $R_v$ value corresponding to the full length of the H chain in the figure.
In general, $C$ is some system-dependent constant as will be shown in the next Section.
A key point is that, for the same $R_o$ but different $h$, the total energy curves are nearly parallel 
across almost the entire range of $R_v$, even before the plateau region;
as $R_o$ is increased, the energy difference between pairs of curves with different $h$
quickly converges.
Thus, relative energies such as binding energies or potential energy curves converge rapidly with increasing localization radii, as we demonstrate next.

\begin{figure} %Fig. 5
\begin{center}
\includegraphics[width=0.45\textwidth]{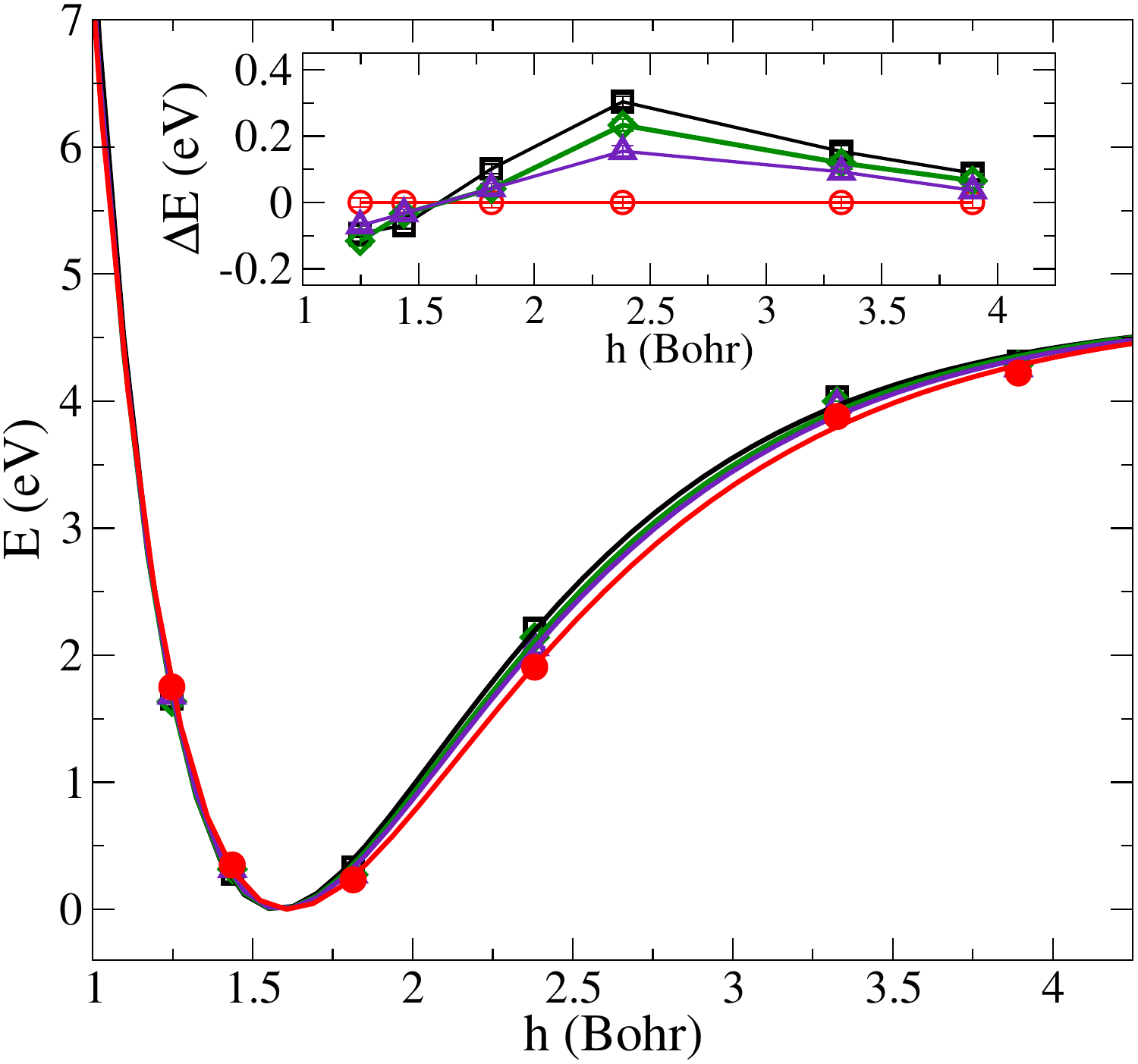}
\caption{\label{O+H20_PEC} 
O/H$_{20}$ PECs vs. the separation, $h$, between O and H$_{20}$ (as per Fig.~\ref{O+H20}).
Filled red circles indicate the PEC for the fully correlated O/H$_{20}$ chain, {\it i.e.} no downfolding or embedding;
black, green, and purple open symbols are for $(R_o,R_v)=(3.7,6.2)$, $(R_o,R_v)=(3.8, 8.3)$, and $(R_o,R_v)=(7.2, 9.8)$, respectively, in Bohr units; the solid lines
are Morse fits. The inset shows deviations compared to the fully correlated PEC (same symbols as in the main figure).
Stochastic error bars are smaller than the symbols.}
\end{center}
\end{figure}
 
The computed potential energy curves (PECs) for the O/H$_{20}$ chain are shown as a function of the O-atom height, $h$, 
in Fig.~\ref{O+H20_PEC}.
Embedding/downfolding PECs are shown for three choices of the cutoff radii and compared to exact AFQMC results, 
({\i.e.}, AFQMC for the entire O/H$_{20}$ chain without downfolding/embedding approximations).
The choice of the ($R_o$, $R_v$) cutoffs was guided 
by the cutoff-radii convergence properties shown in Fig.~\ref{O+H20_EvRvirt}.
Quantitatively good agreement is found across all values of $h$ for these cutoffs.
The smallest cutoffs shown, $(R_o,R_v)=(3.7, 6.2)$ Bohr, correspond to an active region consisting of
the O atom and the 4 nearest H atoms. The corresponding effective Hamiltonian operates on a greatly reduced Hilbert space, 
yielding significant computational savings. 
For ($R_o$, $R_v$)$ = $($7.2$, $9.8$) Bohr, the dimension of the single particle Hilbert space, $M$, is reduced from 113 for the full system (with atomic core frozen), to only 43 and the number of active electrons, $N$, is reduced from 13 to only 6 per spin sector. Based on the theoretical scaling of AFQMC ($\propto M^2N^2$), this reduction in the size of the active Hilbert space corresponds to a  reduction in computational cost by a factor, $f \approx 27$.
For ($R_o$, $R_v$)$ = $($3.8$, $8.3$) Bohr, the effective $M$ = $42$ and $N$ = $5$ giving $f \approx 50$.
($R_o$, $R_v$)$ = $($3.8$, $6.2$) Bohr, has $M$ = $37$ and $N$ = $5$ giving $f \approx 63$.
The reduction in computational cost for \mbox{O/H$_{20}$}, even for the largest choice of cutoff radii, 
 is significant in its own right but is modest when compared to the results of the next section.

\section{Applications}
\label{sec:Application}

In this section our embedding method is applied to carbon chain
systems in a graphitic environment. 
Our choice is motivated by the
recent successful fabrication of long linear carbon chains (LLCCs)
confined within double walled carbon nanotubes (DWCNTs),\cite{Shi2016} and the
possible technological and scientific applications of
TM-functionalized LLCCs.

\subsection{1-Dimensional correlated system : Ti capped linear carbon chain}
\label{sec:corrSys}

We first apply our method to a H-Ti-C$_{29}$-H linear chain. The initial geometry was obtained as follows.
First, an \mbox{H-C$_{30}$-H} chain was fully relaxed with DFT PW91 calculations using NWChem. 
One H-C end-unit was replaced by an H-Ti unit and relaxed, keeping fixed the geometry of the remaining C$_{29}$-H unit. The Ti-C bond is a triple bond. The coordinates of atomic centers are listed
in Table~\ref{sup-tab:H-Ti-C29-H-geom}.
We used  cc-pVDZ basis sets for Ti and C, and the 6-31g basis set
for the terminating H atoms.
Due to near linear dependence of the
basis functions in some of the carbon chain systems, we slightly modified the C cc-pVDZ basis 
by changing some of the p-function exponents on the C atoms (given in Fig.~\ref{sup-bas:custom-cc-pVDZ-basis}). 
The basis sets are given in Fig.~\ref{sup-bas:custom-cc-pVDZ-basis} (used for most Ti-C bond lengths) and in
Fig.~\ref{sup-bas:custom-cc-pVDZ-basis-1.6} (used for the compressed bond length of 3.0 Bohr).
Although accurate treatment of transition metal atoms in many-body calculations usually requires 
larger basis sets and multi-determinant trial wave functions~\cite{Purwanto2016},
we focus here on the systematics of the
downfolding and embedding method, and
 restrict ourselves to the modest Ti cc-pVDZ basis 
and single-determinant trial wave functions.
(We do, however, examine basis size-dependence below for a shorter H-Ti-C$_{7}$-H linear chain.)

\begin{figure} %Fig. 6
\begin{center}
\includegraphics[width=0.45\textwidth]{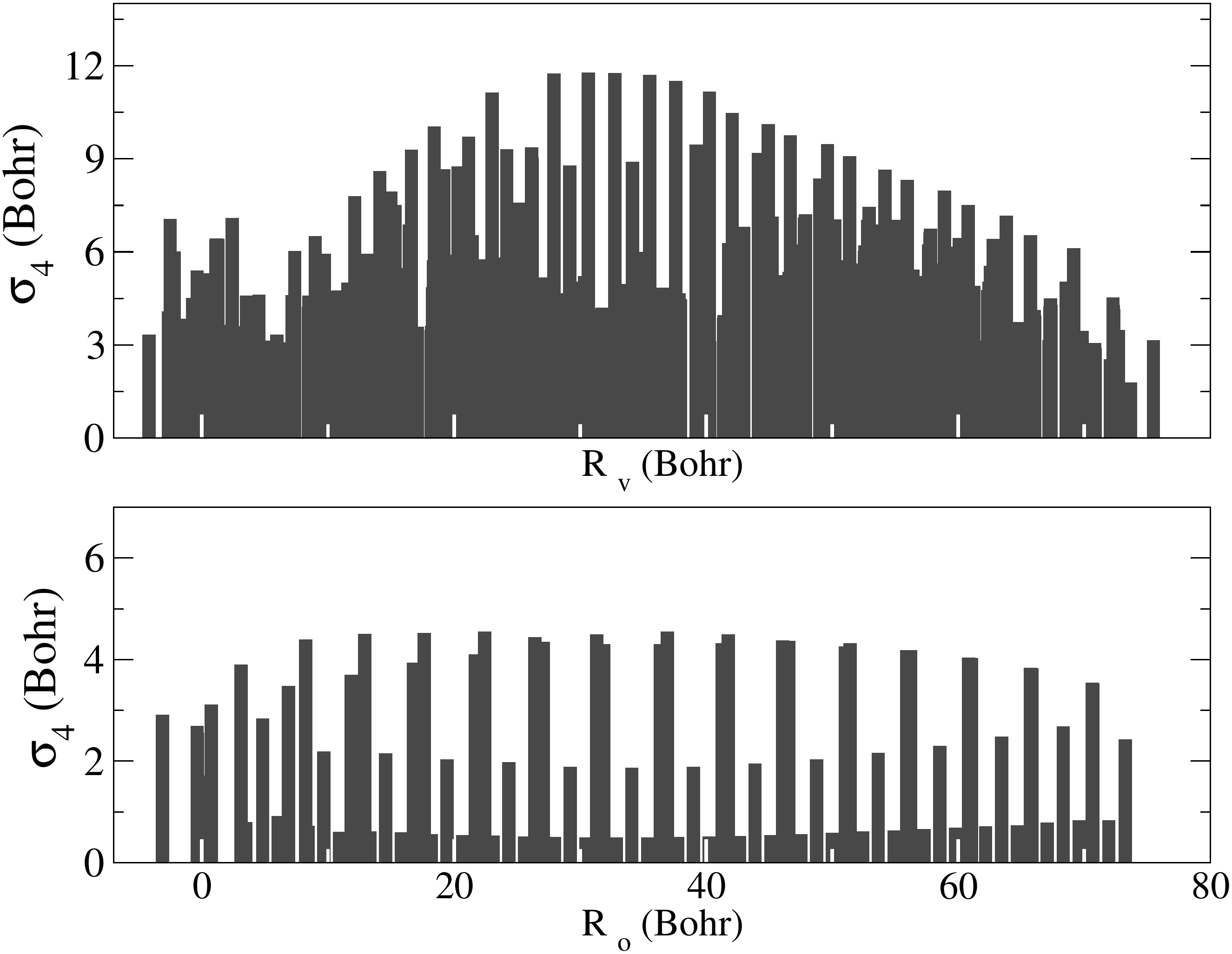}
\caption{
\label{fig:H-Ti-C29-H-centroids4thMom} H-Ti-C$_{29}$-H  
bar plot of the fourth central moment $\sigma_4$ [Eq.~(\ref{eq:pthMomSpread})] of the localized virtual (upper panel) and localized occupied (lower panel) orbitals 
vs. the orbital centroid position along the chain.  
The origin is set to the position of the Ti atom.
}
\end{center}
\end{figure}

Figure~\ref{fig:H-Ti-C29-H-centroids4thMom} shows bar plots
 of the fourth central moment orbital spreads, $\sigma_4$, 
of the localized virtual (upper panel) and localized occupied (lower panel) orbitals.
Since FM2 localization of the virtuals did not converge, FB localization was used for the occupied and FM localization was used for the virtual orbitals.
Although many virtual states in Fig.~\ref{fig:H-Ti-C29-H-centroids4thMom} have rather 
large $\sigma_4$ values, most of these orbitals are far from the Ti atom. 
At the same time, occupied and virtual orbitals near Ti are seen to have much smaller $\sigma_4$ than most other orbitals. 
It is therefore expected that convergence with respect to the localization radii, $R_o$ and $R_v$, will only weakly 
depend on $\sigma_4$ of the least local orbital. This is confirmed below. 

\begin{figure} %Fig. 7
\begin{center}
\includegraphics[width=0.45\textwidth]{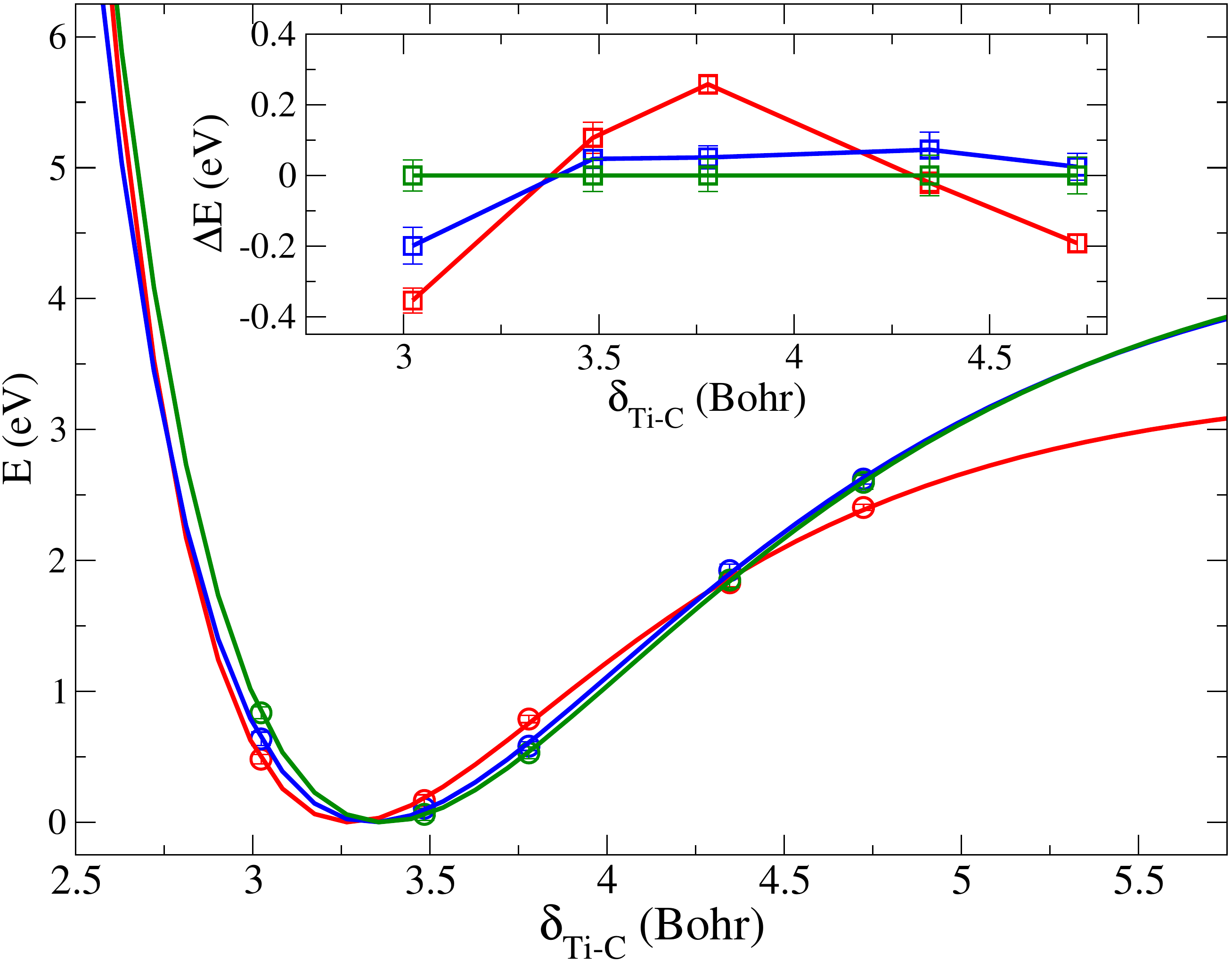}
\caption{
\label{fig:H-Ti-C29-H-PEC} 
Main panel: H-Ti-C$_{29}$-H PECs vs. Ti-C bond length $\delta_\textrm{Ti-C}$. 
Red, blue and green correspond, respectively, to cutoff ($R_o$, $R_v$) values of (4.9, 8.9), (9.7, 13.7), and (14.7, 18.6) in Bohr units.
Solid lines are Morse fits. An energy offset was applied to align the minima of the Morse fits.
Inset: PEC deviations from the (14.7, 18.6) reference.
}
\end{center}
\end{figure}

\begin{figure} %Fig. 8
\begin{center}
\includegraphics[width=0.45\textwidth]{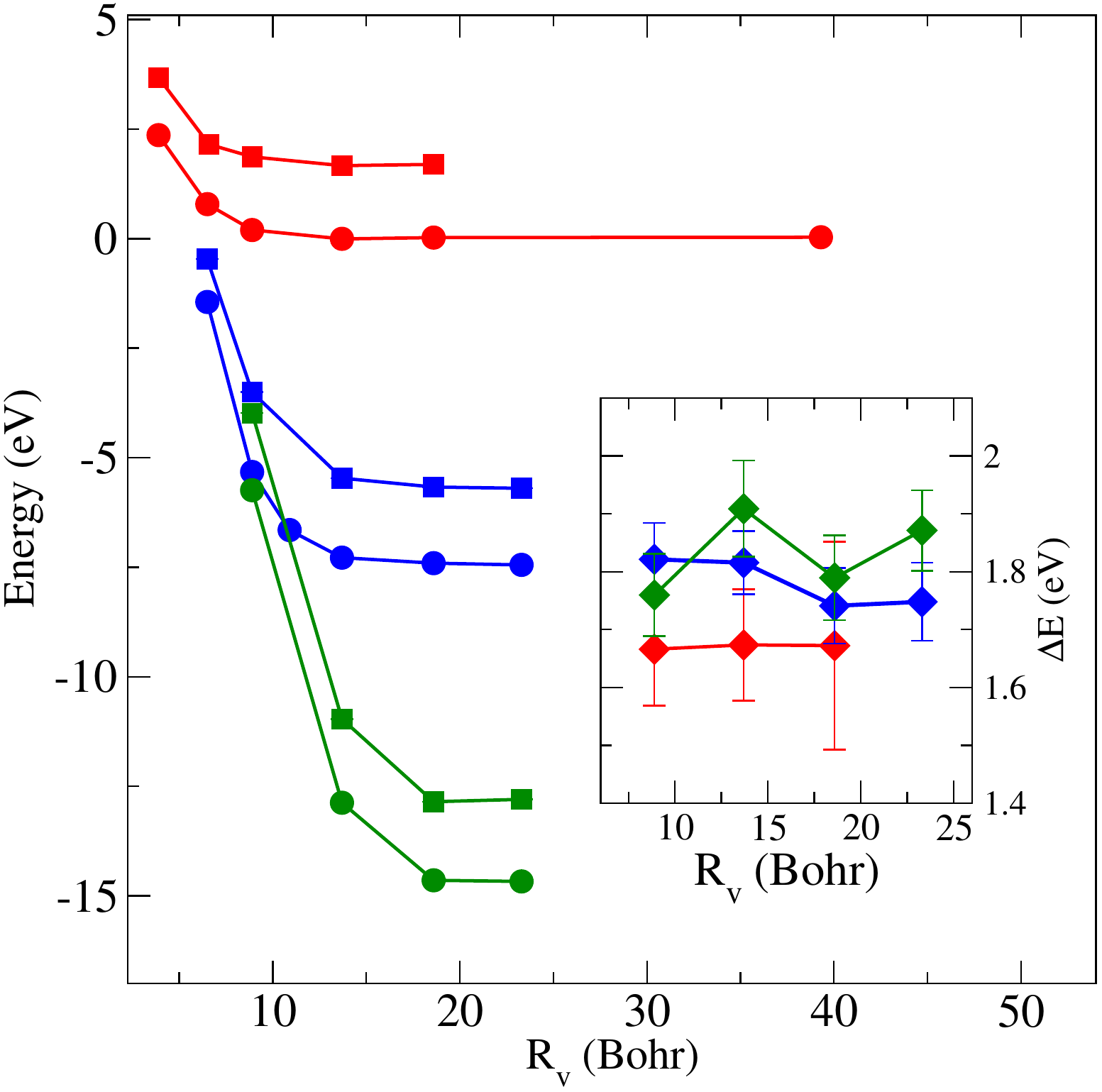}
\caption{
\label{fig:H-Ti-C29-H-EvsRvirt} 
H-Ti-C$_{29}$-H AFQMC energy vs. $R_{v}$.
 Red corresponds to $R_o=4.9$ Bohr, blue corresponds to $R_o=9.7$ Bohr and green corresponds to $R_o=14.7$ Bohr. Filled circles represent the equilibrium $\delta_\textrm{Ti-C}=3.40$ Bohr and filled squares represent a stretched $\delta_\textrm{Ti-C}=4.35$ Bohr. Solid lines are a guide to the eye. Stochastic error bars are smaller than the symbols.
All AFQMC total energies were shifted by the same constant.
In the inset, the energy difference, $\Delta E = E(4.35$ Bohr$) - E(3.40$ Bohr$)$, is plotted against $R_v$ at fixed $R_o$ (same color coding as in the main figure).}
\end{center}
\end{figure}

The H-Ti-C$_{29}$-H PECs are shown in Fig.~\ref{fig:H-Ti-C29-H-PEC} for three sets of ($R_o$, $R_v$).
For each $R_o$ in Fig.~\ref{fig:H-Ti-C29-H-PEC},
Fig.~\ref{fig:H-Ti-C29-H-EvsRvirt} shows the convergence for fixed $R_o$ as a function of $R_v$, 
for two Ti-C bond lengths ($\delta_\textrm{Ti-C}=3.40$ and 4.35 Bohr). 
The energy in Fig.~\ref{fig:H-Ti-C29-H-EvsRvirt} is seen to
quickly plateau as $R_v$ becomes greater than $R_o$. 
Thus, the $R_o$ cutoff effectively sets the size of the number of one-particle states that must be correlated in the effective Hamiltonian.
The rapid convergence with increasing numbers of virtual orbitals  is similar for curves 
with the same $R_o$ but different $\delta_\textrm{Ti-C}$.  
The curves for $\delta_\textrm{Ti-C}=3.40$ and 4.35 Bohr 
are nearly parallel across almost the entire range of $R_v$, reflecting good
cancellation of errors for the Ti-C chain, similar to O/H$_{20}$ (Fig.~\ref{O+H20_EvRvirt}).
This is shown by the inset of Fig.~\ref{fig:H-Ti-C29-H-EvsRvirt}.
The PECs in Fig.~\ref{fig:H-Ti-C29-H-PEC} are converged in $R_o$ at about $R_o$ = 9.7 Bohr for all bond lengths except $\delta_\textrm{Ti-C} = 3.0$ Bohr.

\subsection{Finite-Size and Basis-Set Effects in the Linear Chain}
\label {sec:BasisSize}

In the next subsection, we present embedding/downfolding calculations for a H-Ti-C$_n$-H chain on a graphitic substrate.
These were done for a shorter \mbox{H-Ti-C$_{7}$-H} chain. 
Calculations for shorter chains facilitate direct comparisons with full AFQMC calculations, allowing us to focus on 
the embedding/downfolding approximations, while reducing the additional computational cost due to the graphitic substrates.
Starting from the equilibrium structure  \mbox{H-Ti-C$_{29}$-H} chain,
22 interior C atoms were removed, keeping the remaining bond lengths the same. Atomic positions are given in Table~\ref{sup-tab:H-Ti-C7-H-geom}.
Unless otherwise stated, the custom C cc-pVDZ basis used to remove near linear dependence in \mbox{H-Ti-C$_{29}$-H} (Fig.~\ref{sup-bas:custom-cc-pVDZ-basis}) is used for \mbox{H-Ti-C$_{7}$-H} as well.
FB localization is used for both the occupied and virtual orbitals as FM did not converge for \mbox{H-Ti-C$_{7}$-H} on the graphitic substrate studied in the next section.
 
\begin{figure} %fig. 9
\begin{center}
\includegraphics[width=0.45\textwidth]{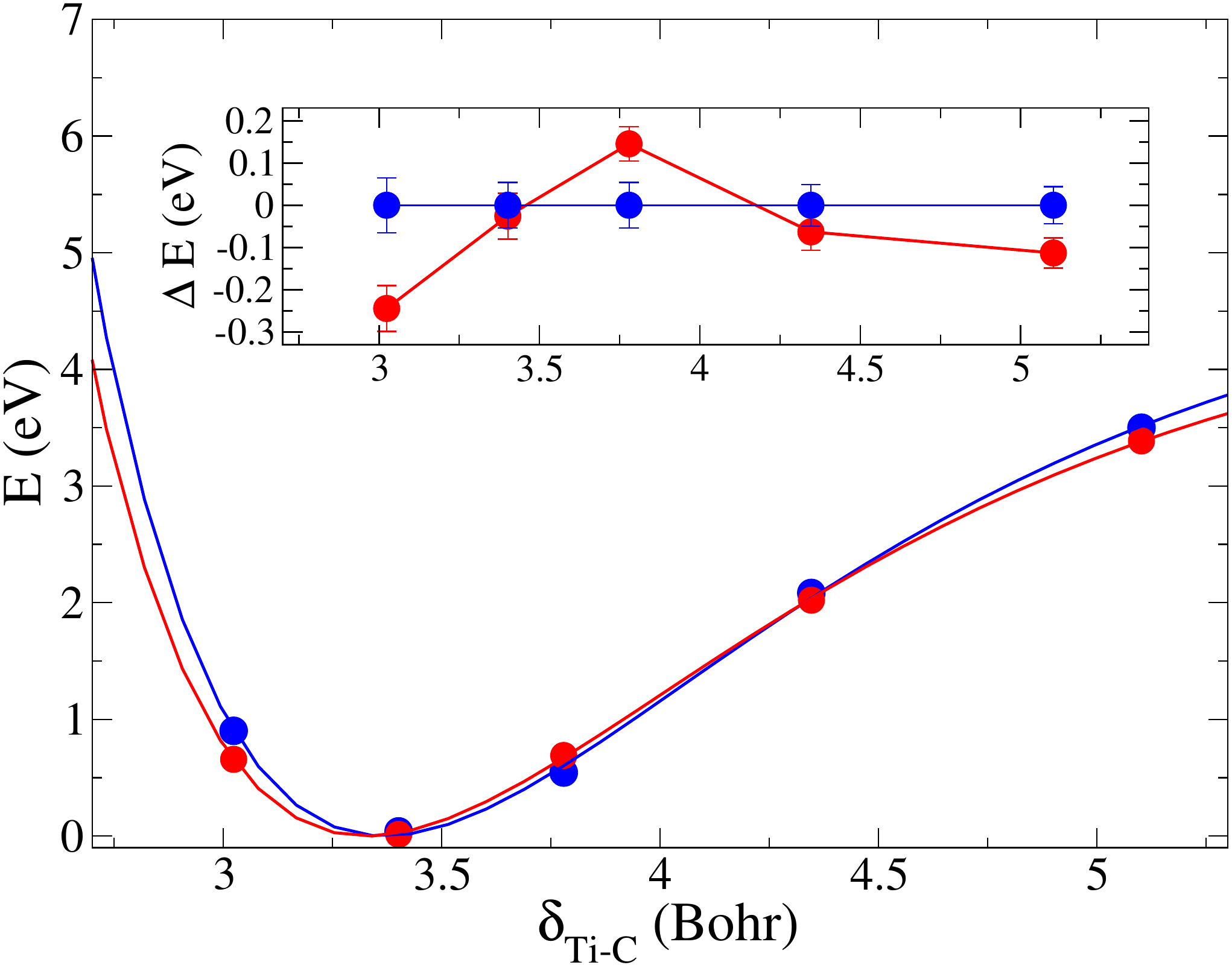}
\caption{
\label{fig:TiC7-PEC-full} 
Main panel: H-Ti-C$_{7}$-H PECs vs. Ti-C bond length $\delta_\textrm{Ti-C}$ using both embedding/downfolding with ($R_o$, $R_v$) $=$ (9.7, 13.7) Bohr (red circles), 
and fully correlated calculation (blue circles). Solid lines are Morse fits.
An energy offset was applied to align the Morse fit minima.
Stochastic error bars are smaller than the symbols. 
Inset: Difference between the embedding/downfolding PEC and
the fully correlated calculation. 
 The line is a guide to the eye.
 }
\end{center}
\end{figure}

\begin{figure} % Fig. 10
\begin{center}
\includegraphics[width=0.45\textwidth]{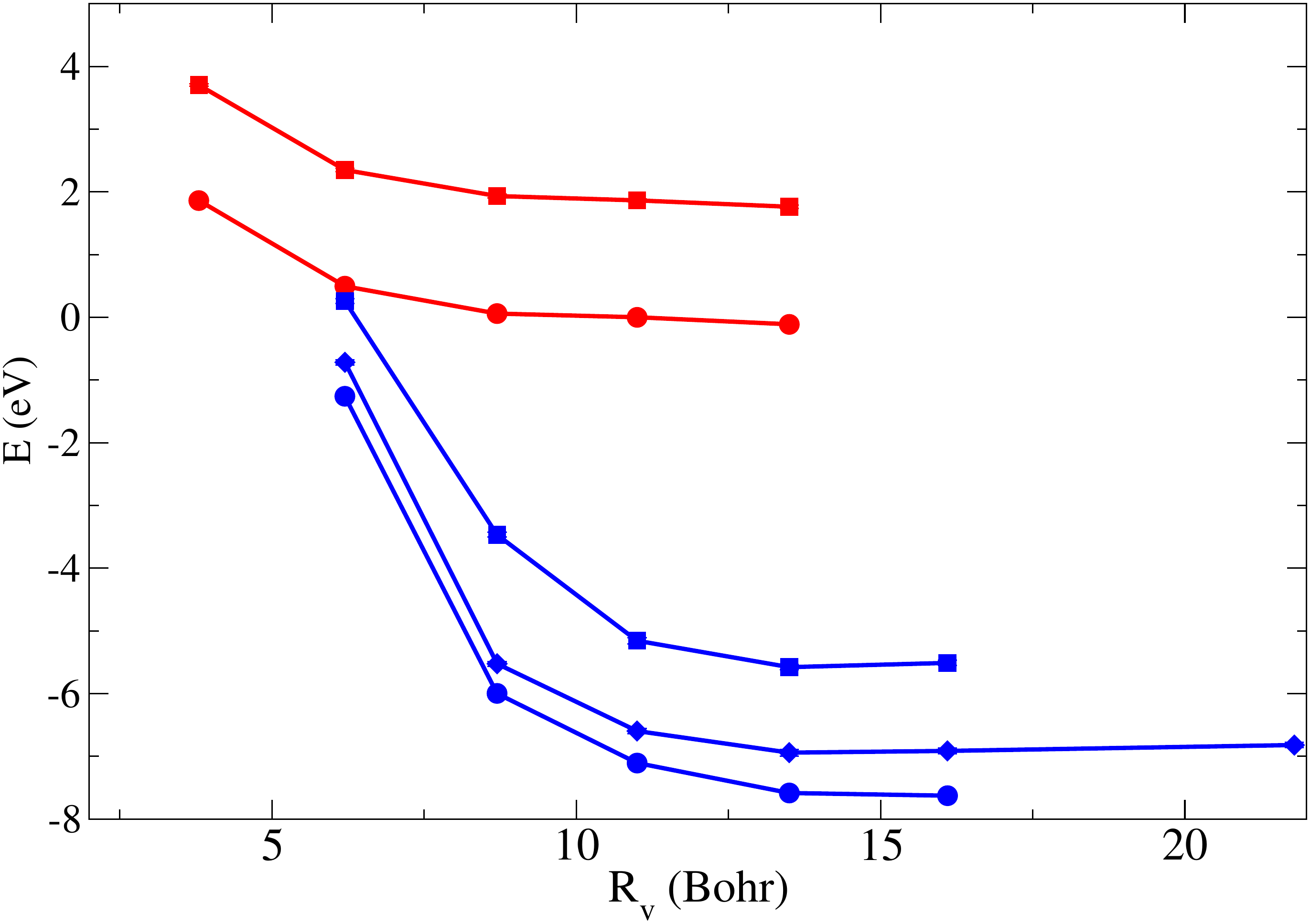}
\caption{
\label{fig:H-Ti-C7-H-EvsRvirt} 
H-Ti-C$_{7}$-H AFQMC energy vs. $R_{v}$ for two values of $R_o$. Red and blue correspond respectively, 
to $R_o=4.9$ and  $R_o=9.7$ Bohr.  Filled circles represent the equilibrium $\delta_\textrm{Ti-C}=3.40$ Bohr and filled squares represent a stretched $\delta_\textrm{Ti-C}=4.35$ Bohr. For $R_o=9.7$ Bohr, results are also shown 
for $\delta_\textrm{Ti-C}=3.02$ Bohr, the compressed bond.
Solid lines are a guide to the eye. 
All AFQMC total energies were shifted by the same constant.
}
\end{center}
\end{figure}
\begin{figure*} %Fig. 11
\begin{center}
\includegraphics[width=0.9\textwidth]{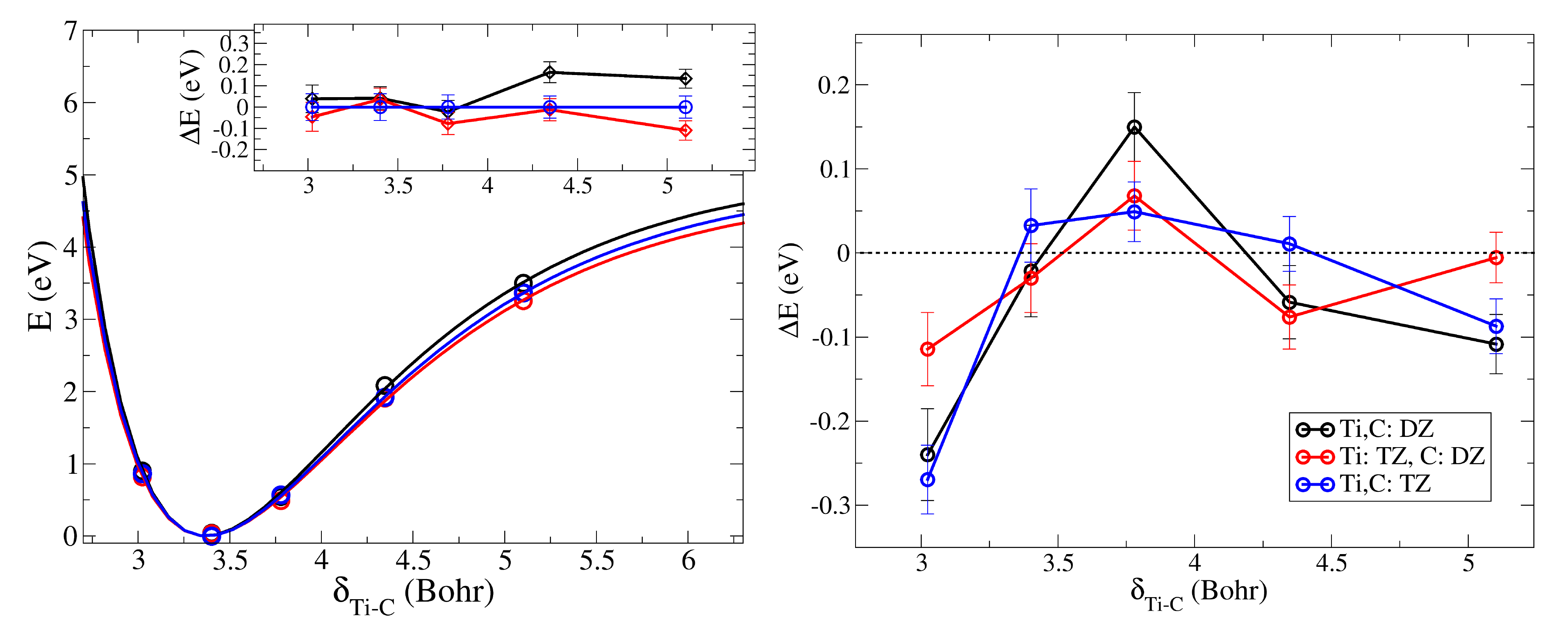}
\caption{\label{fig:embedding-bas}
Basis set effects for fully correlated calculations (left panel) and embedding/downfolding calculations (right panel).
Three basis sets are compared
(see legend in right panel).
The left panel shows the fully-correlated PECs and Morse fits. The inset of the left panel plots the energy difference with respect to cc-pVTZ for both Ti and C.
The right panel plots the energy difference between the embedding/downfolding result and the fully correlated result for each basis set considered.
The error bars (all panels) reflect the stochastic uncertainty of the individual AFQMC energies (as opposed to the joint stochastic uncertainty).}
\end{center}
\end{figure*}

We first calculated energies for \mbox{H-Ti-C$_{7}$-H}  as a function of the Ti-C bondlength $\delta_\textrm{Ti-C}$.
Figure~\ref{fig:TiC7-PEC-full} compares the \mbox{H-Ti-C$_{7}$-H} embedding PEC to that of a fully correlated AFQMC calculation without
embedding or downfolding approximations; 
the inset of the figure shows that deviations from full AFQMC are less than about 0.1 eV, except at the shortest 
$\delta_\textrm{Ti-C}$, where it is $\simeq 0.24$~eV, and at an intermediate bondlength, $\delta_\textrm{Ti-C} = 3.78$ Bohr, where it is $\simeq 0.15$ eV. 
The error bars in the inset reflect the stochastic error of the individual energy measurements (as opposed to the joint stochastic error).
Fig.~\ref{fig:H-Ti-C7-H-EvsRvirt} shows the \mbox{H-Ti-C$_{7}$-H} energy vs. $R_v$ for two fixed values of $R_o$ and a few
values of $\delta_\textrm{Ti-C}$. The convergence in $R_v$ for fixed $R_o$ is similar to the longer H-Ti-C$_{29}$-H chain (compare to Fig.~\ref{fig:H-Ti-C29-H-EvsRvirt}) for $\delta_\textrm{Ti-C} = 3.40$ and $4.35$ Bohr.
However, for $\delta_\textrm{Ti-C} = 3.0$ Bohr, using $R_o = 9.7$ Bohr, the convergence in $R_v$ differs from the other two $\delta_\textrm{Ti-C}$ plotted. This is reflected in the inset of Fig.~\ref{fig:TiC7-PEC-full} in which
it is observed that the greatest deviation from the fully correlated result occurs at this particular value for $\delta_\textrm{Ti-C}$ when using ($R_o$, $R_v$) = (9.7, 13.7) Bohr.
It seems that for a slightly larger $R_v$ of 16.1 Bohr that the relative energy between $\delta_\textrm{Ti-C} = 3.0$ Bohr and other $\delta_\textrm{Ti-C}$ may be more converged.

%\begin{figure*} %Fig. 11
%\begin{center}
%\includegraphics[width=0.9\textwidth]{figs/figure11.pdf}
%\caption{\label{fig:embedding-bas}
%Basis set effects for fully correlated calculations (left panel) and embedding/downfolding calculations (right panel).
%Three basis sets are compared
%(see legend in right panel).
%The left panel shows the fully-correlated PECs and Morse fits. The inset of the left panel plots the energy difference with respect to cc-pVTZ for both Ti and C.
%The right panel plots the energy difference between the embedding/downfolding result and the fully correlated result for each basis set considered.
%The error bars (all panels) reflect the stochastic uncertainty of the individual AFQMC energies (as opposed to the joint stochastic uncertainty).}
%\end{center}
%\end{figure*}
%
The left panel of Fig.~\ref{fig:embedding-bas} shows the basis set dependence of the fully correlated AFQMC PECs.
The cc-pVTZ basis used for C atoms was modified similarly to the C cc-pVDZ basis (Fig.~\ref{sup-bas:custom-cc-pVTZ-basis}).
As seen in the figure, all basis sets give results within about 0.05 eV near equilibrium. 
Only small basis-set effects are found for the fully correlated treatment.
In the right panel of Fig.~\ref{fig:embedding-bas}, the deviation of embedded/downfolded results from the fully correlated results are plotted (similar to the inset of the left panel).
Again, only small basis-set effects are observed and the
embedding and downfolding effects are seen to be largely independent of basis set zeta-quality.

\begin{figure} %Fig. 12
\begin{center}
\includegraphics[width=0.45\textwidth]{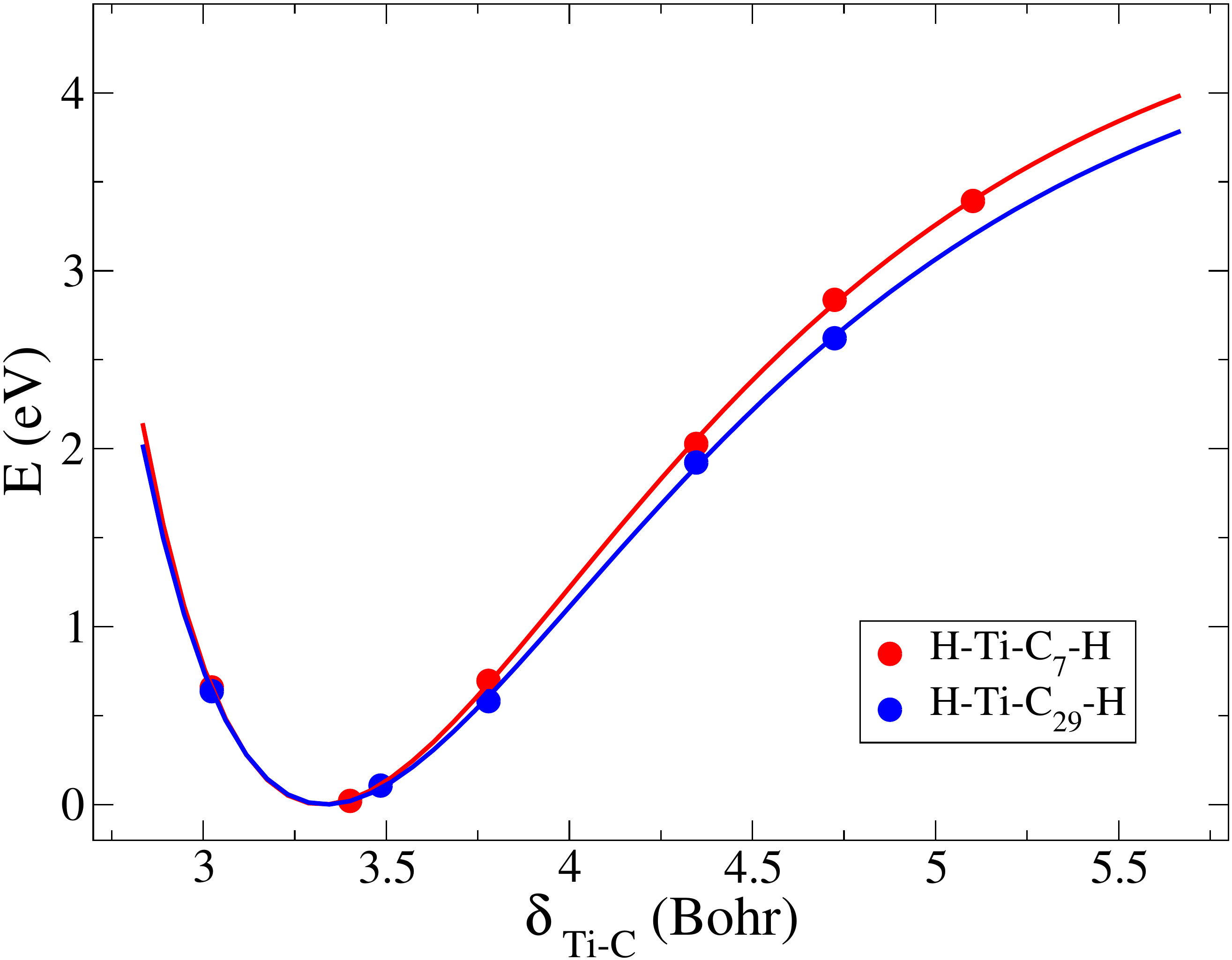}
\caption{
\label{fig:H-Ti-Cn-H-PEC} H-Ti-C$_{n}$-H PECs vs. Ti-C bond length $\delta_\textrm{Ti-C}$. 
Localization cutoffs were the same for both chains, $R_o$ and $R_v$ equal to 9.7 and 13.7 Bohr, respectively.
An energy offset was applied to align the minima of the Morse fits (solid lines).
 Stochastic error bars are smaller than the symbols.
}
\end{center}
\end{figure}
Results for the \mbox{H-Ti-C$_{7}$-H} and  \mbox{H-Ti-C$_{29}$-H} chains
are compared in Fig.~\ref{fig:H-Ti-Cn-H-PEC}. 
The custom cc-pVDZ basis set given in Fig.~\ref{sup-bas:custom-cc-pVDZ-basis-1.6} is used for \mbox{H-Ti-C$_7$-H} (only at $\delta_\textrm{Ti-C} = 3.0$ Bohr) to facilitate comparison to \mbox{H-Ti-C$_{29}$-H} at that particular bond length.
Cutoff radii for both systems were chosen as $R_o= 9.7$ and $R_v = 13.7$ Bohr.
As indicated by Figs.~\ref{fig:H-Ti-C29-H-EvsRvirt} and \ref{fig:H-Ti-C7-H-EvsRvirt}, this choice of cutoffs yields well converged results with the possible exception of the compressed bond length, $\delta_\textrm{Ti-C} = 3.0$ Bohr.
The PECs of both systems are in good qualitative agreement. We note that the potential well is slightly deeper for \mbox{H-Ti-C$_{7}$-H} than for \mbox{H-Ti-C$_{29}$-H} despite the fact 
that the active subsystems, at the AFQMC level of theory, are nearly identical in both systems. We discuss this point further in Section~\ref{sec:Discussion}.

\subsection{Interaction of the Linear Chain with a Graphitic Substrate}
\label{sec:envInt}

\begin{figure} %Fig. 13
\begin{center}
\includegraphics[width=0.45\textwidth]{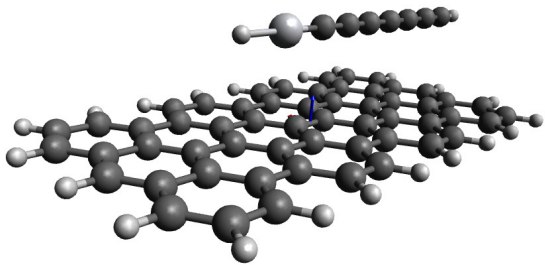}
\caption{
\label{fig:TiC7atGrapheneGeom} 
\mbox{H-Ti-C$_{7}$-H / H$_{22}$C$_{56}$} geometry. 
C and H atoms are depicted as dark gray and small light-gray spheres, respectively; 
the Ti atom is indicated by the large light-gray sphere. 
The \mbox{H-Ti-C$_7$-H} chain is in the reflection symmetry plane of H$_{22}$C$_{56}$.
As $\delta_\textrm{Ti-C}$ is varied, the other atoms are kept fixed, 
and the C atom farthest from Ti within the chain is always aligned above a C atom 
on the zig-zag edge of the H$_{22}$C$_{56}$ ribbon. The chain height above the substrate is fixed at  5.99 Bohr.
The lateral dimensions of the ribbon are $14.0 \times 29.9$ Bohr, excluding H terminations.
The atomic coordinates are given in Table~\ref{sup-tab:H-Ti-C7-HatH22-C56-geom}.
The molecular visualization was generated using Avogadro~\cite{Hanwell2012}.
}
\end{center}
\end{figure}

There has been recent interest in studying the role of environmental interactions on carbon chain systems~\cite{Wanko2016}. Recent experiments have
observed long linear carbon chains (LLCCs) confined within double walled carbon nanotubes (DWCNTs). \cite{Shi2016} 
Here we consider a simpler model of carbon chains near a graphitic planar substrate. 
Given the similar properties of the Ti-capped C$_{29}$ and C$_7$ chains in the previous section, we study  \mbox{H-Ti-C$_{7}$-H}
chains on a finite-size graphitic ribbon, \mbox{H$_{22}$C$_{56}$}, as depicted in Fig.~\ref{fig:TiC7atGrapheneGeom}, where
the dangling bonds of the boundary C atoms are saturated by H atoms.  
The C-C bond lengths are fixed to the experimentally observed value in graphene $\delta_\textrm{C-C}=2.68$ Bohr; the
C-H bond lengths are set to 2.06 Bohr. 
The lateral dimensions of the ribbon are $14.0 \times 29.9$ Bohr, excluding H terminations.
The chain is placed parallel to the ribbon at a vdW height of 5.99 Bohr, taken from a carbon chain in Ref.~\citenum{Wanko2016},
and within the reflection symmetry plane of the ribbon. 
For the equilibrium $\delta_\textrm{Ti-C}$ found above for \mbox{H-Ti-C$_{7}$-H} in vacuum, we found only small 
residual atomic forces $\simeq 0.2$~eV/Bohr using vdW-including DFT/PBE-D2 NWCHEM calculations.
This indicates that the system is close to its equilibrium geometry for the vacuum-relaxed  \mbox{H-Ti-C$_{7}$-H} chain geometry.
The positions of all atoms, including both the \mbox{H-Ti-C$_7$-H} chain atoms and the \mbox{H$_{22}$C$_{56}$} graphitic substrate atoms, are given in Table~\ref{sup-tab:H-Ti-C7-HatH22-C56-geom}.

RHF calculations for \mbox{H-Ti-C$_{7}$-H}/\mbox{H$_{22}$C$_{56}$} 
were done as follows.  The atoms in the  \mbox{H-Ti-C$_{7}$-H} chain used the same basis sets as in vacuum (Fig.~\ref{sup-bas:custom-cc-pVDZ-basis}). 
For the \mbox{H$_{22}$C$_{56}$} substrate, however, we excluded d-functions for the C atoms 
to reduce the computational cost (Fig.~\ref{sup-bas:custom-cc-pVDZ-basis-substrate}).
FB localization was used for the RHF occupied sector as in the other calculations. 
FM localization for the virtual orbitals did not converge in ERKALE, however, so
FB localization of the virtuals was used instead. 
Although the substrate FB virtuals have somewhat large $\sigma_4$ values, they are well separated
from the chain FB virtuals.
Moreover, in \mbox{H-Ti-C$_{7}$-H}/\mbox{H$_{22}$C$_{56}$}, 
the chain localized orbitals closely resemble those of the vacuum H-Ti-C$_{7}$-H chain, both in their centroid positions and in
their $\sigma_4$ values. 
It was therefore straightforward to
 select $R_o$ and $R_v$ cutoffs within the chain (similar to those used in vacuum) 
and separate cutoffs to select the active region within the substrate.

\begin{figure*} % Fig. 14
\begin{center}
\includegraphics[width=0.9\textwidth]{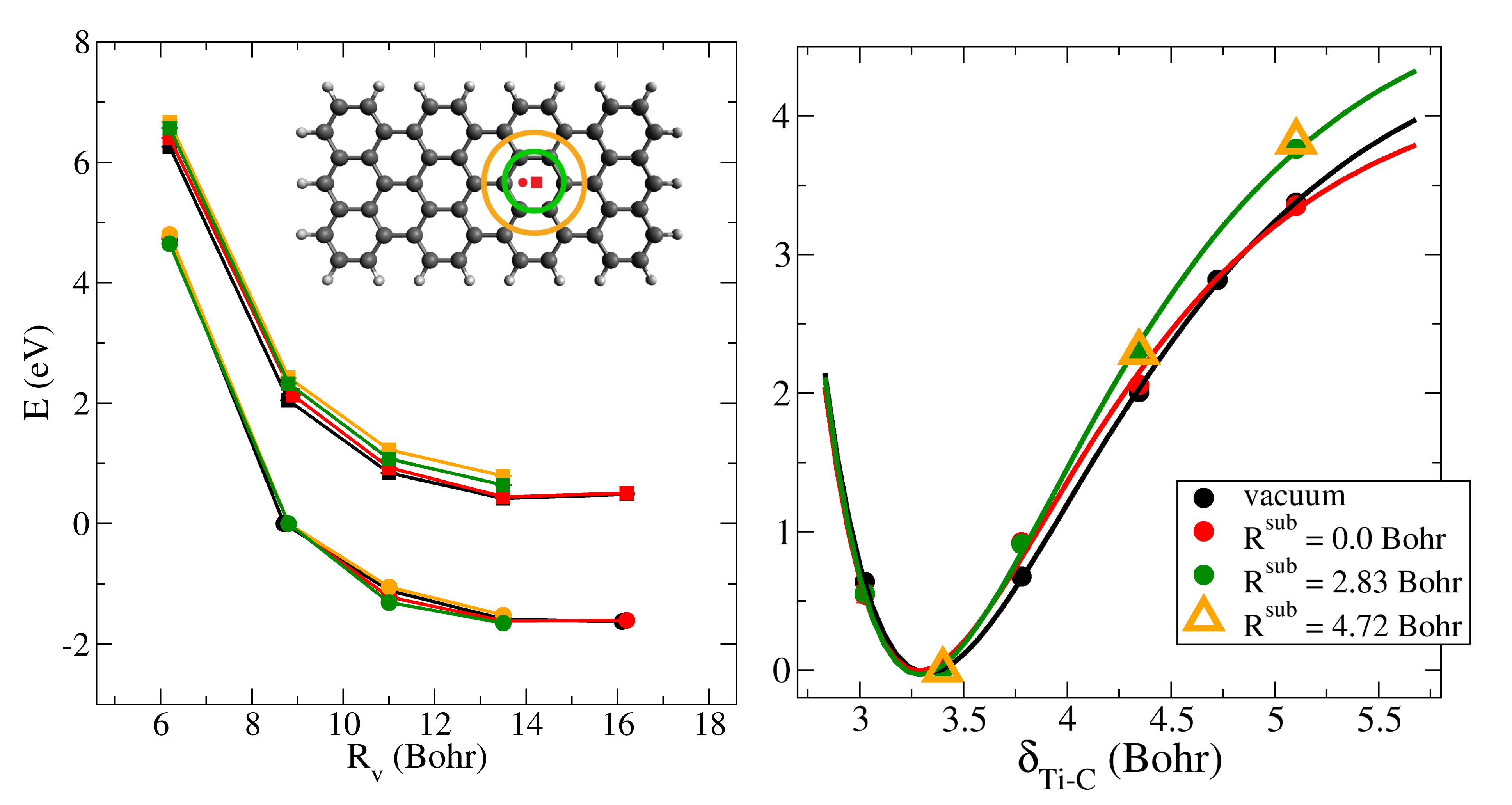}
\caption{
\label{fig:TiC7atGrapheneE} 
\mbox{H-Ti-C$_7$-H / H$_{22}$C$_{56}$}  \mbox{energy vs. $R_{v}$} (left panel) and PECs
as a function of $\delta_\textrm{Ti-C}$ (right panel)
for $R^{sub}=0$ (red),  2.83 (green), and 4.72 (orange) Bohr. 
The latter two are illustrated in the inset of the left panel.
For comparison, results for \mbox{H-Ti-C$_7$-H} in vacuum are plotted with black symbols and lines.
In the left panel, all the calculations are for $R_{o} = 9.7$~Bohr, and for two Ti-C bond lengths,
$\delta_\textrm{Ti-C}=3.40$ (near equilibrium; circles) and 4.35~Bohr (squares).
As indicated in the inset, the Ti atom is centered on the 6-fold hollow site when $\delta_\textrm{Ti-C}=4.35$~Bohr (molecular visualization generated using Avogadro~\cite{Hanwell2012}).
The $R^{sub}=0$ curve corresponds to the frozen-orbital embedding approximation for the occupied ribbon states (see text), 
{\i.e.}, the only active-space states are localized on the \mbox{H-Ti-C$_7$-H} chain.
In the right panel, all computations use $(R_o, R_v) = (9.7,13.7)$ Bohr. 
Morse fits are shown as solid lines, except for \mbox{$R^{sub} = 4.72$ Bohr},  where only two bond lengths are shown.
Stochastic error bars are smaller than the symbols. 
The curves are aligned to have the same energy minima.
}
\end{center}
\end{figure*}

The \mbox{H-Ti-C$_7$-H} chain components of the active space are defined using similar in-chain radial cutoffs, $R_o$ and $R_v$, as in the previous section.
If we required a similar criterion for the ribbon, we would define the in-plane active region
using substrate radii, \mbox{$R_o^{sub}=\sqrt{R_o^2-h^2}$} and \mbox{$R_v^{sub}=\sqrt{R_v^2-h^2}$}, where $h=5.99$ Bohr is the
chain height above the ribbon, and where $R_o^{sub}$ and $R_v^{sub}$ are measured from the projected position of the Ti atom. 
In this case, occupied (virtual) orbitals in the ribbon would only be correlated if $R_o > h \simeq 6$ Bohr ($R_v > h \simeq 6$ Bohr). 
Thus, for $R_o=4.9$ Bohr in Fig.~\ref{fig:H-Ti-C7-H-EvsRvirt}, all the ribbon occupied states would be frozen.
For the choice $(R_o, R_v) = (9.7,13.7)$ Bohr in Fig.~\ref{fig:TiC7-PEC-full}, 
this would yield \mbox{$R_o^{sub}=7.6$} and \mbox{$R_v^{sub}=12.3$} Bohr.
The latter choice would result in larger calculations than we wanted to undertake in this paper. 
Instead, calculations were done setting $R_o^{sub}=R_v^{sub}=R^{sub}$ , where $R^{sub}$ is a common radius 
measured from the 6-fold hollow site. Calculations were done for three values of 
$R^{sub}=0$, 2.83 and 4.72 Bohr, where the latter two are illustrated by the circles in the inset of Fig.~\ref{fig:TiC7atGrapheneE} (left panel).
The $R^{sub-nu}=0$ case corresponds to the
embedding approximation for all of the occupied ribbon states; 
all the virtual ribbon orbitals are excluded from the active space. Thus for $R^{sub}=0$, chain-ribbon
interactions are present only in the 1-body contributions to the
effective Hamiltonian ($\hat{V}^{\mathbb{I} - \mathbb{A}}$ in Eq.~\ref{eq:Hdownfold}). 

The left panel of Fig.~\ref{fig:TiC7atGrapheneE} shows calculated AFQMC energies of \mbox{H-Ti-C$_7$-H / H$_{22}$C$_{56}$} 
for three values of $R^{sub}$, as a function of $R_{v}$ and for two Ti-C bond lengths. 
All the curves are for $R_{o}=9.7$~Bohr (the same $R_o$ that was 
used for the H-Ti-C$_7$-H chain in vacuum in Figs.~\ref{fig:embedding-bas}, and~\ref{fig:H-Ti-Cn-H-PEC}).
For comparison, the analogous \mbox{H-Ti-C$_7$-H}  vacuum results from Fig.~\ref{fig:H-Ti-C7-H-EvsRvirt} are also shown.
All the curves, for the same $\delta_\textrm{Ti-C}$, are seen to be within $\simeq 0.40$ eV of each other.
The $R^{sub}=0$ and vacuum \mbox{H-Ti-C$_7$-H} curves are nearly identical, 
which shows that the 1-body embedding contributions for $R^{sub}=0$ are
very small. This is not too surprising given the large chain-ribbon separation of 5.99 Bohr.
The  $\delta_\textrm{Ti-C}=3.40$ and 4.35~Bohr curves are nearly parallel to each other and show similar convergence with increasing $R_v$,
leading to rapid convergence of energy differences. 
The similar convergence of relative energies between correlated substrate treatments and the H-Ti-C$_7$-H chain in vacuum indicates that
converged in-chain $R_o$ and $R_v$ can be chosen by considering only the chain system in vacuum. 
This would be considerably less expensive than performing many trial calculations in the presence of a larger environmental system.

The \mbox{H-Ti-C$_7$-H / H$_{22}$C$_{56}$} PECs corresponding to the left panel of Fig.~\ref{fig:TiC7atGrapheneE}  are shown
in the right panel of the same figure.
For stretched $\delta_\textrm{Ti-C}$, the \mbox{$R^{sub}=2.83$ and 4.72 Bohr} PECs (with active states in the substrate) are about
$\simeq 0.2-0.4$ eV larger than the $R^{sub}=0$ (frozen substrate)
and vacuum-chain \mbox{H-Ti-C$_7$-H} PECs.
The correlated-substrate well depths are therefore deeper than those for the vacuum or frozen substrate case. 
The frozen substrate and the vacuum chain PECs are in very good agreement except at 3.78 Bohr.
At that specific bond length, the RHF solution landscape is particularly complicated and so it is possible that the RHF trial wavefunction used for \mbox{H-Ti-C$_{7}$-H} in vacuum is slightly different than 
the analogous RHF trial wavfunction used for \mbox{H-Ti-C$_{7}$-H / H$_{22}$C$_{56}$} in terms of the single-particle states corresponding to the chain.
This conclusion is further supported by the very strong agreement between the frozen substrate and correlated substrate treatments at that particular $\delta_\textrm{Ti-C}$.
We note that the PECs for the $R^{sub}=2.83$ and 4.72 Bohr treatments 
are in excellent agreement for all $\delta_\textrm{Ti-C}$ computed, suggesting that convergence in $R^{sub}$ has been achieved.
Further calculations should be done, however, to check the convergence with increasing $R^{sub}$ to confirm this.
Additionally, the effect of separate substrate occupied and virtual localization radii, $R_o^{sub}$ and $R_v^{sub}$, should be considered.
Ribbon finite-size effects should also be examined using larger (especially longer) ribbons.

\section{Discussion}
\label{sec:Discussion}

As shown in Sections~\ref{sec:modelSys} and~\ref{sec:Application}, embeddding/downfolding 
energies converge rapidly with increasing cutoff radius $R_v$ for fixed $R_o$ (Figs.~\ref{O+H20_EvRvirt}, \ref{fig:H-Ti-C29-H-EvsRvirt}, \ref{fig:H-Ti-C7-H-EvsRvirt}, and the left panel of Fig.~\ref{fig:TiC7atGrapheneE}).
Thus, $R_v$ can be chosen naturally as $R_v = R_o + C$, where \mbox{$C \simeq 2 - 6$ Bohr} is a system-dependent constant.
The convergence rate, in terms of $C$, is observed to be nearly the same for closely related systems.
For example, all of the \mbox{H-Ti-C$_n$-H} systems studied in Sec.~\ref{sec:Application} have $C \approx 4$ Bohr,
including \mbox{H-Ti-C$_{7}$-H / C$_{56}$H$_{22}$} (independent of the in-plane cutoff, $R^{sub}$) as can be seen in Figs.~\ref{fig:H-Ti-C29-H-EvsRvirt},~\ref{fig:H-Ti-C7-H-EvsRvirt},~\ref{fig:TiC7atGrapheneE}~(left panel).
Convergence is even more rapid for relative energies (inset of Fig.~\ref{fig:H-Ti-C29-H-EvsRvirt}), due to cancellation of errors.
In the case of \mbox{O/H$_{20}$}, the most converged PEC in Fig.~\ref{O+H20_PEC}, with ($R_o$, $R_v$) = (7.8, 9.8) Bohr, agrees with the full AFQMC treatement to within 0.1 eV for almost all O-atom heights.
For \mbox{H-Ti-C$_{29}$-H},
the inset of Fig.~\ref{fig:H-Ti-C29-H-PEC} shows that the embedding results are nearly converged;
the \mbox{($R_o$, $R_v$) = (9.7, 13.7) Bohr} PEC deviates from the  \mbox{($R_o$, $R_v$) = (14.7, 18.6) Bohr} PEC by no more than 0.1 eV for all bond lengths
except for the most compressed bond length where the deviation is about 0.2 eV. 
Similar behavior is observed for \mbox{H-Ti-C$_{7}$-H} in Fig.~\ref{fig:TiC7-PEC-full}
with a maximum deviation of about 0.24 eV, with most bond lengths deviating by 0.1 eV or less.
Thus, local embedding/downfolding can be applied with little loss of quantitative accuracy. 

Using the embedding/downfolding effective Hamiltonian significantly reduces the computational cost compared to the AFQMC treatment of the full Hamiltonian.
For the \mbox{H-Ti-C$_{29}$-H} chain system studied in Sec.~\ref{sec:corrSys} 
using ($R_o$, $R_v$) $=$ ($9,7$, $13.7$) Bohr, which was found to be converged for most $\delta_\textrm{Ti-C}$, 
the number of electrons per spin, N, is reduced from 65 (core states frozen) to 13, and the number of basis set functions, M, is reduced from 419 to only 100, 
reducing the computational cost by a factor of about 440, with AFQMC scaling as $\propto M^2N^2$. 
For \mbox{H-Ti-C$_{7}$-H /H$_{22}$C$_{56}$} studied in Section~\ref{sec:envInt}, the full system
has $M$=625 and $N$ (per spin) = 144.
Using localization radii ($R_o$, $R_v$) $=$ ($9,7$, $13.7$) Bohr and allowing the 6-fold hollow site nearest to the \mbox{Ti} atom to be correlated ($R^{sub} = 2.83$ Bohr), $M$ becomes 130 and $N$ becomes 22, reducing the cost by nearly 3 orders of magnitude (a factor of 990).
With such reductions, we were able to treat, with AFQMC, the \mbox{H-Ti-C$_{29}$-H} and the \mbox{H-Ti-C$_{7}$-H /H$_{22}$C$_{56}$} systems using only modest computing resources. Thus, the effective system size which can be treated is greatly extended by the local embedding and effective downfolding approximation.
We note that these computations were performed using the modest cc-pVDZ basis. 
For larger basis sets, the cost
may be expected to be reduced by a similar factor, since the fraction of basis functions which are assigned to $\mathbb{A}$ will be roughly the same as for smaller basis sets, assuming a similar distribution of localized single-particle orbitals in both basis sets.

In Section~\ref{sec:BasisSize}, basis-convergence effects on the local embedding and downfolding approximation were considered. For fixed ($R_o$, $R_v$), the deviation of AFQMC results using embedding and downfolding from the full AFQMC treatment seems to be largely independent of basis set quality as seen in the right panel of Fig.~\ref{fig:embedding-bas}.
This can be exploited in order to study convergence  in the localization radii while using a smaller, and therefore less expensive, basis set before using higher quality basis sets to obtain quantitatively accurate results.

Finite-size effects were studied in Section~\ref{sec:BasisSize}, comparing the PECs for \mbox{H-Ti-C$_{7}$-H} and \mbox{H-Ti-C$_{29}$-H} (Fig.~\ref{fig:H-Ti-Cn-H-PEC}).
Good qualitative agreement was observed, with slight quantitative differences in the depth of the potential well.
The geometry of  \mbox{H-Ti-C$_{7}$-H} is based on the geometry of  \mbox{H-Ti-C$_{29}$-H} with the chain truncated after 7 C atoms and with a H termination applied.
As can be seen from the distribution and spatial extent of local orbitals in both systems (Fig.~\ref{sup-fig:H-Ti-Cn-H-centHist}), the active regions have spatially similar localized occupied and virtual orbitals.
As a result, the active regions of both chains are the same spatial size and have the same number of electrons and have basis set sizes that differ by only one.
The inactive host regions differ, of course, containing 4 and 26 carbon atoms in the \mbox{H-Ti-C$_{7}$-H} and \mbox{H-Ti-C$_{29}$-H} chains respectively.
Therefore the effective Hamiltonians of the two chains differ mostly
in their mean-field embedding contributions, {\it i.e.}, in $\Vop^{\mathbb{I} - \mathbb{A}}$ from Eq.~\ref{eq:Hdownfold}.
Unitarily localized Kohn-Sham orbitals, obtained from DFT, or natural orbitals, obtained using some many-body method could lead to different one-body size effects in $\Vop^{\mathbb{I} - \mathbb{A}}$ than the orbitals used here.

Orbital localization using the FM2 cost function with Hessian-based trust-region minimization is regarded as the method of choice for
obtaining well-localized virtual orbitals.\cite{Hoyvik2012-TR} 
As mentioned, we used the ERKALE program, which uses gradient-based minimization, to localize both occupied and virtual orbitals in the present work.
The consequence of using gradient-based methods to minimize the FM2 cost function is 
possible non-convergence, as is the case for the systems studied here.
FM did not converge for \mbox{H-Ti-C$_{7}$-H /H$_{22}$C$_{56}$} and so FB, which is expected to produce virtual orbitals with long tails, 
was used for all \mbox{H-Ti-C$_{7}$-H} systems. 
Despite the use of such virtual orbitals,
 rapid convergence in $R_v$ at fixed $R_o$ is observed. 
This suggests that local embedding/downfolding 
for a single active cluster is robust against basis sets which have not been optimally localized.
Further study is needed to determine the effects of the system dimensionality on the convergence in $R_v$.
In this regard, the one-dimensional geometry of the chain systems is certainly advantageous.
It is also observed in all \mbox{H-Ti-C$_{n}$-H} systems studied here that the least local orbitals tend to be localized far from the \mbox{Ti-C} cluster  (see Figs.~\ref{fig:H-Ti-C29-H-centroids4thMom} and ~\ref{sup-fig:H-Ti-Cn-H-centHist}), providing some additional explanation for the rapid convergence observed in these systems.
Using FM2 virtual orbitals is likely to be especially efficient for two- and three-dimensional system geometries.

\section{Summary}
\label{sec:Summary}

In summary, a local embedding and effective downfolding approximation
has been developed and implemented in which a local cluster is explicitly
treated at the AFQMC level of theory.
Previously, only local embedding of the occupied sector had been used in AFQMC, which requires the large virtual sector to be explicitly treated in all AFQMC computations.
Here, we have developed a local effective downfolding scheme for the virtual sector and combined it with the local embedding approximation.
The computational effort of performing AFQMC computations using embedding/downfolding
is greatly reduced (by as much as 3 orders of magnitude in the examples studied here) 
 from the full AFQMC treatment.
Therefore, the effective system size which can be feasibly treated with AFQMC is significantly increased. 
We have studied the convergence systematics in the localization radii, $R_o$ and $R_v$, which define the embedding and downfolding regions. 
Relative energies computed using local embedding and effective downfolding are observed to converge rapidly to the full AFQMC result.
Furthermore, the rate of convergence is seen to be similar for closely related systems such as the \mbox{H-Ti-C$_n$-H} systems studied in this paper.
The convergence is seen to be largely independent of the zeta-quality of the basis set used.

We have used only RHF trial wave functions, $\ket{\Psi_T}$, to study the convergence systematics in the cutoff radii, $R_o$ and $R_v$.
In some cases open-shell or multi-determinant $\ket{\Psi_T}$ may be required to achieve high accuracy AFQMC results.
The embedding/downfolding approximation can be generalized to these cases. 
For example, in multi-determinant $\ket{\Psi_T}$'s, the determinants are typically expressed in terms of a single set of orthonormal canonical orbitals,
which can then be localized by the same unitary transformation.
Furthermore, it is expected that the systematics in $R_o$ and $R_v$ will be similar in these cases since the action of the embedding and downfolding transformation
merely reduces the effective size of the single-particle Hilbert space which is spanned by a set of unitarily localized orbitals.
Further study is necessary to determine the systematics for more general $\ket{\Psi_T}$.

\acknowledgements
This work is supported by DOE (Grant No. DE-SC0001303), ONR (Contract No. N00014-17-1-2237), and the Simons Foundation.
This work was performed using computing facilities at William \& Mary which were provided by contributions from the National Science Foundation,
 the Commonwealth of Virginia Equipment Trust Fund and the Office of Naval Research.
The Flatiron institute is a division of the Simons Foundation.

\bibstyle{natbib}
\bibliography{article-main.bib}

\end{document}

%\tableofcontents
%\def\table{\def\figurename{Table}\figure}
%\let\endtable\endfigure 
%\renewcommand\listfigurename{List of Figures and Tables}
\clearpage
\listoftables

\listoffigures

\clearpage

\setcounter{equation}{0}
\setcounter{figure}{0}
\setcounter{table}{0}
\setcounter{page}{1}
\makeatletter
\renewcommand{\theequation}{S\arabic{equation}}
\renewcommand{\thefigure}{S\arabic{figure}}
\renewcommand{\thetable}{S\arabic{table}}
\renewcommand{\bibnumfmt}[1]{[S#1]}
\renewcommand{\citenumfont}[1]{S#1}

\begin{table}[htb!]
\centering
\begin{tabular}{ l l l l }
\hline
Atom & x (\r{A}) & y(\r{A}) & z(\r{A}) \\
\hline
O  &  0.000000  &  0.0  & $h$ \\
H  &  -0.4704240745  &  0.0  &  0.0 \\
H  &  0.4704240745  &  0.0  &  0.0 \\
H  &  1.4112722235  &  0.0  &  0.0 \\
H  &  2.3521203725  &  0.0  &  0.0 \\
H  &  3.2929685215  &  0.0  &  0.0 \\
H  &  4.2338166705  &  0.0  &  0.0 \\
H  &  5.1746648195  &  0.0  &  0.0 \\
H  &  6.1155129685  &  0.0  &  0.0 \\
H  &  7.0563611175  &  0.0  &  0.0 \\
H  &  7.9972092665  &  0.0  &  0.0 \\
H  &  8.9380574155  &  0.0  &  0.0 \\
H  &  9.8789055645  &  0.0  &  0.0 \\
H  &  10.8197537135  &  0.0  &  0.0 \\
H  &  11.7606018625  &  0.0  &  0.0 \\
H  &  12.7014500115  &  0.0  &  0.0 \\
H  &  13.6422981605  &  0.0  &  0.0 \\
H  &  14.5831463095  &  0.0  &  0.0 \\
H  &  15.5239944585  &  0.0  &  0.0 \\
H  &  16.4648426075  &  0.0  &  0.0 \\
H  &  17.4056907565  &  0.0  &  0.0 \\
\end{tabular}
\caption{\label{tab:O+H20-geom} \mbox{O/H$_{20}$} atomic positions.
%The H-H bond lengths are set to the equilibrium value of a symmetrically H-chain.
The \mbox{H-H} bond length is fixed at 0.94 \r{A} (1.78 Bohr), corresponding approximately to the equilibrium bond length of the H chain.
The variable O atom separation from the \mbox{H$_{20}$} chain is indicated as $h$.}
\end{table}

\begin{table}[htb!]
\centering
\begin{tabular}{ l l l l }
\hline
Atom & x (\r{A}) & y(\r{A}) & z(\r{A}) \\
\hline
H  &  0.00000000  &  0.00000000  &  -19.61807759  \\
Ti  &  0.00000000  &  0.00000000  &  -17.75883316  \\
C  &  0.00000000  &  0.00000000  &  -15.91476478  \\
C  &  0.00000000  &  0.00000000  &  -14.56680791  \\
C  &  0.00000000  &  0.00000000  &  -13.31403993  \\
C  &  0.00000000  &  0.00000000  &  -11.98246807  \\
C  &  0.00000000  &  0.00000000  &  -10.72221809  \\
C  &  0.00000000  &  0.00000000  &  -9.39740173  \\
C  &  0.00000000  &  0.00000000  &  -8.13331512  \\
C  &  0.00000000  &  0.00000000  &  -6.81207565 \\
C  &  0.00000000  &  0.00000000  &  -5.54575283 \\
C  &  0.00000000  &  0.00000000  &  -4.22660585 \\
C  &  0.00000000  &  0.00000000  &  -2.95897701  \\
C  &  0.00000000  &  0.00000000  &  -1.64101239  \\
C  &  0.00000000  &  0.00000000  &  -0.37268881  \\
C  &  0.00000000  &  0.00000000  &  0.94471639  \\
C  &  0.00000000  &  0.00000000  &  2.21325368  \\
C  &  0.00000000  &  0.00000000  &  3.53065888  \\
C  &  0.00000000  &  0.00000000  &  4.79898256  \\
C  &  0.00000000  &  0.00000000  &  6.11694723  \\
C  &  0.00000000  &  0.00000000  &  7.38457612  \\
C  &  0.00000000  &  0.00000000  &  8.70372309  \\
C  &  0.00000000  &  0.00000000  &  9.97004587  \\
C  &  0.00000000  &  0.00000000  &  11.29128534  \\
C  &  0.00000000  &  0.00000000  &  12.55537190  \\
C  &  0.00000000  &  0.00000000  &  13.88018838  \\
C  &  0.00000000  &  0.00000000  &  15.14043841  \\
C  &  0.00000000  &  0.00000000  &  16.47201027  \\
C  &  0.00000000  &  0.00000000  &  17.72477839 \\
C  &  0.00000000  &  0.00000000  &  19.07273559  \\
C  &  0.00000000  &  0.00000000  &  20.30828945  \\
H  &  0.00000000  &  0.00000000  &  21.38632828  \\
\end{tabular}
\caption{\label{tab:H-Ti-C29-H-geom} \mbox{H-Ti-C$_{29}$-H} atomic positions. An \mbox{H-C$_{30}$-H} chain was fully relaxed using DFT/pw91 while constrained to remain linear by imposing c2v symmetry. The end \mbox{H-C} unit was removed and replaced by an \mbox{H-Ti} unit. The \mbox{H-Ti} unit was then relaxed using DFT/pw91 while holding all other atomic positons fixed, again using c2v symmetry to enforce a linear geometry. For the AFQMC calculations, the \mbox{Ti-C} bond length was manually set to the specified value while holding the \mbox{H-Ti} separation fixed. The \mbox{Ti-C} bond length given in the present table is the relaxed \mbox{Ti-C} bond length obtained as described. Both geometry relaxations were performed using NWChem in a standard cc-pVDZ basis for Ti and C and in a 6-31g basis for H.}
\end{table}

\begin{table}[htb!]
\centering
\begin{tabular}{ l l l l }
\hline
Atom & x (\r{A}) & y(\r{A}) & z(\r{A}) \\
\hline
H  &  0.0  &  3.54258827  &  3.17 \\
Ti  &  0.0  &  1.68334384  &  3.17 \\
C  &  0.0  &  -0.11665616  &  3.17 \\
C  &  0.0  &  -1.46461303  &  3.17 \\
C  &  0.0  &  -2.71738101  &  3.17 \\
C  &  0.0  &  -4.04895287  &  3.17 \\
C  &  0.0  &  -5.30920285  &  3.17 \\
C  &  0.0  &  -6.63401921  &  3.17 \\
C  &  0.0  &  -7.89810582  &  3.17 \\
H  &  0.0  &  -8.97614465  &  3.17 \\
\end{tabular}
\caption{\label{tab:H-Ti-C7-H-geom} \mbox{H-Ti-C$_7$-H} atomic positions. Geometry is based on the \mbox{H-Ti-C$_{29}$-H} chain system in Table~\ref{tab:H-Ti-C29-H-geom}. \mbox{H-Ti-C$_{29}$-H} is truncated after the first 7 atoms in the chain system. A terminating H atom is placed at the same C-H bond length as in the \mbox{H-Ti-C$_{29}$-H} chain system. The \mbox{H-Ti-C$_7$-H} chain has also been rotated %by an angle $-\frac{\pi}{2}$ about the x-axis 
and translated. % to a different origin.
}
\end{table}

\begin{table}[htb!]\scriptsize
%\centering
\begin{minipage}[l]{0.45\textwidth}
\begin{tabular}{ l l l l }
\hline
Atom & x (\r{A}) & y(\r{A}) & z(\r{A}) \\
\hline
H  &  0.0  &  3.54258827  &  3.17 \\
Ti  &  0.0  &  1.68334384  &  3.17 \\
C  &  0.0  &  -0.11665616  &  3.17 \\
C  &  0.0  &  -1.46461303  &  3.17 \\
C  &  0.0  &  -2.71738101  &  3.17 \\
C  &  0.0  &  -4.04895287  &  3.17 \\
C  &  0.0  &  -5.30920285  &  3.17 \\
C  &  0.0  &  -6.63401921  &  3.17 \\
C  &  0.0  &  -7.89810582  &  3.17 \\
H  &  0.0  &  -8.97614465  &  3.17 \\
H  &  -4.63069619  &  -7.77831758  &  0.00000000 \\
H  &  -4.66422620  &  -5.31424460  &  0.00000000 \\
H  &  -4.65930659  &  -3.37178298  &  0.00000000 \\
H  &  -4.65898750  &  -0.95997161  &  0.00000000 \\
H  &  -4.65898750  &  0.95997161  &  0.00000000 \\
H  &  -4.65930659  &  3.37178298  &  0.00000000 \\
H  &  -4.66422620  &  5.31424460  &  0.00000000 \\
H  &  -4.63069619  &  7.77831758  &  0.00000000 \\
C2  &  -3.67492009  &  -7.22847115  &  0.00000000 \\
C2  &  -3.68874897  &  -5.82047350  &  0.00000000 \\
C2  &  -2.49202984  &  -5.06419666  &  0.00000000 \\
C2  &  -2.48526224  &  -3.60546160  &  0.00000000 \\
C2  &  -3.68806218  &  -2.85768374  &  0.00000000 \\
C2  &  -3.68757329  &  -1.47275005  &  0.00000000 \\
C2  &  -2.48112828  &  -0.72579472  &  0.00000000 \\
C2  &  -2.48112828  &  0.72579472  &  0.00000000 \\
C2  &  -3.68757329  &  1.47275005  &  0.00000000 \\
C2  &  -3.68806218  &  2.85768374  &  0.00000000 \\
C2  &  -2.48526224  &  3.60546160  &  0.00000000 \\
C2  &  -2.49202984  &  5.06419666  &  0.00000000 \\
C2  &  -3.68874897  &  5.82047350  &  0.00000000 \\
C2  &  -3.67492009  &  7.22847115  &  0.00000000 \\
C2  &  -2.46783161  &  7.92144925  &  0.00000000 \\
H  &  -2.44446065  &  9.02395382  &  0.00000000 \\
C2  &  -1.23170743  &  -7.21162185  &  0.00000000 \\
C2  &  -1.23970331  &  -5.76334748  &  0.00000000 \\
H  &  -2.44446065  &  -9.02395382  &  0.00000000 \\
C2  &  -2.46783161  &  -7.92144925  &  0.00000000 \\
C2  &  0.00000000  &  -5.04400380  &  0.00000000 \\
C2  &  0.00000000  &  -3.60450400  &  0.00000000 \\
C2  &  -1.24291895  &  -2.88744021  &  0.00000000 \\
C2  &  -1.24169534  &  -1.44244200  &  0.00000000 \\
C2  &  0.00000000  &  -0.71848972  &  0.00000000 \\
C2  &  0.00000000  &  0.71848972  &  0.00000000 \\
\end{tabular}
\end{minipage}
\begin{minipage}[l]{0.45\textwidth}
\begin{tabular}{ l l l l }
C2\hphantom{At}  &  -1.24169534  &  1.44244200  &  0.00000000 \\
C2  &  -1.24291895  &  2.88744021  &  0.00000000 \\
C2  &  0.00000000  &  3.60450400  &  0.00000000 \\
C2  &  0.00000000  &  5.04400380  &  0.00000000 \\
C2  &  -1.23970331  &  5.76334748  &  0.00000000 \\
C2  &  -1.23170743  &  7.21162185  &  0.00000000 \\
C2  &  0.00000000  &  7.89810582  &  0.00000000 \\
H  &  0.00000000  &  9.00128634  &  0.00000000 \\
C2  &  1.23170743  &  -7.21162185  &  0.00000000 \\
C2  &  1.23970331  &  -5.76334748  &  0.00000000 \\
H  &  0.00000000  &  -9.00128634  &  0.00000000 \\
C2  &  0.00000000  &  -7.89810582  &  0.00000000 \\
C2  &  2.49202984  &  -5.06419666  &  0.00000000 \\
C2  &  2.48526224  &  -3.60546160  &  0.00000000 \\
C2  &  1.24291895  &  -2.88744021  &  0.00000000 \\
C2  &  1.24169534  &  -1.44244200  &  0.00000000 \\
C2  &  2.48112828  &  -0.72579472  &  0.00000000 \\
C2  &  2.48112828  &  0.72579472  &  0.00000000 \\
C2  &  1.24169534  &  1.44244200  &  0.00000000 \\
C2  &  1.24291895  &  2.88744021  &  0.00000000 \\
C2  &  2.48526224  &  3.60546160  &  0.00000000 \\
C2  &  2.49202984  &  5.06419666  &  0.00000000 \\
C2  &  1.23970331  &  5.76334748  &  0.00000000 \\
C2  &  1.23170743  &  7.21162185  &  0.00000000 \\
C2  &  2.46783161  &  7.92144925  &  0.00000000 \\
H  &  2.44446065  &  9.02395382  &  0.00000000 \\
C2  &  3.67492009  &  -7.22847115  &  0.00000000 \\
C2  &  3.68874897  &  -5.82047350  &  0.00000000 \\
H  &  2.44446065  &  -9.02395382  &  0.00000000 \\
C2  &  2.46783161  &  -7.92144925  &  0.00000000 \\
C2  &  3.68806218  &  -2.85768374  &  0.00000000 \\
C2  &  3.68757329  &  -1.47275005  &  0.00000000 \\
C2  &  3.68757329  &  1.47275005  &  0.00000000 \\
C2  &  3.68806218  &  2.85768374  &  0.00000000 \\
C2  &  3.68874897  &  5.82047350  &  0.00000000 \\
C2  &  3.67492009  &  7.22847115  &  0.00000000 \\
H  &  4.63069619  &  -7.77831758  &  0.00000000 \\
H  &  4.66422620  &  -5.31424460  &  0.00000000 \\
H  &  4.65930659  &  -3.37178298  &  0.00000000 \\
H  &  4.65898750  &  -0.95997161  &  0.00000000 \\
H  &  4.65898750  &  0.95997161  &  0.00000000 \\
H  &  4.65930659  &  3.37178298  &  0.00000000 \\
H  &  4.66422620  &  5.31424460  &  0.00000000 \\
H  &  4.63069619  &  7.77831758  &  0.00000000 \\
\end{tabular}
\end{minipage}
\caption{\label{tab:H-Ti-C7-HatH22-C56-geom} \mbox{H-Ti-C$_7$-H/H$_{22}$C$_{56}$} atomic positions. The \mbox{C-C} bond lengths are set to the experimentally observed value in graphene of 1.42 \r{A} (2.68 Bohr) and the \mbox{C-H} bond lengths are set to 1.09 \r{A} (2.06 Bohr). We note that two difference Carbon basis sets are used. Atoms specified with atomic symbol ``C'' use the basis given in Fig.~\ref{bas:custom-cc-pVDZ-basis} and atoms specified with atomic symbol ``C2'' use the basis given in Fig.~\ref{bas:custom-cc-pVDZ-basis-substrate}. The geometry of \mbox{H-Ti-C$_7$-H} is unchanged from the geometry used in vacuum given in Table \ref{tab:H-Ti-C7-H-geom}.}
\end{table}

\clearpage

%%%%%%%%%%%%%%%%%%%%%%%%%%%%%%%%%%%%%%%%%%%%%
				    %Begin Figures section %
%%%%%%%%%%%%%%%%%%%%%%%%%%%%%%%%%%%%%%%%%%%%%

\begin{figure}
\begin{center}
\includegraphics[width=0.6\textwidth]{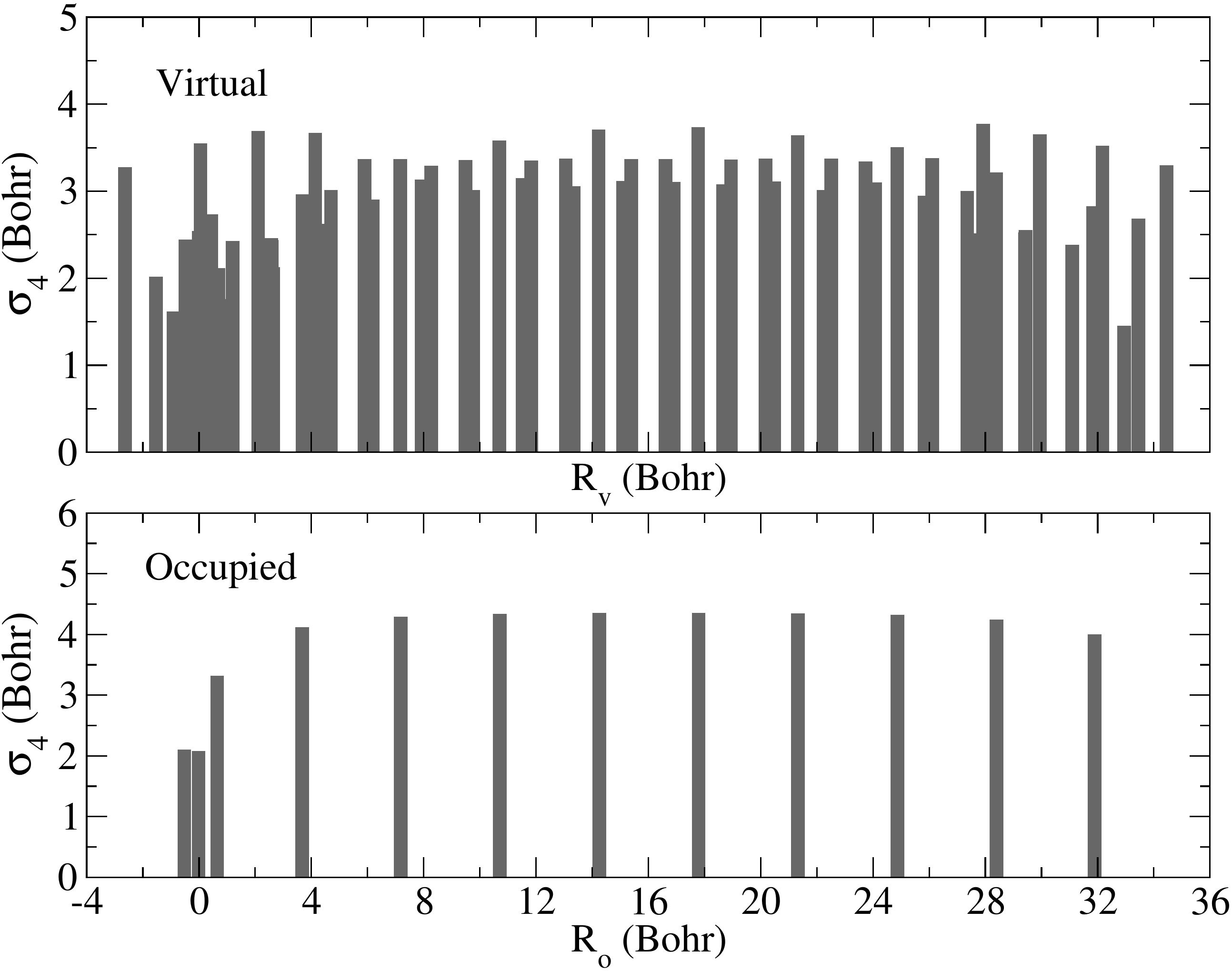}
\caption{
\label{fig:O+H20-centroids4thMom} \mbox{O/H$_{20}$}  
bar plot of the fourth central moment $\sigma_4$ [Eq.~(\ref{eq:pthMomSpread})] of the localized virtual (upper panel) and occupied (lower panel) orbitals 
vs. the orbital centroid position along the H chain.  
The origin is set to the position of the O atom projected onto the H chain.
%\SZ{SZ: Figs 4 and 5 aree not referred to in the text? Consider combining them into one figure, for example similar to Fig 3}
}
\end{center}
\end{figure}

\clearpage

\begin{figure}[htb!]
\begin{tabular}{l l l l }
C&    S& & \\
  & 6665.0000000       &       0.0006920   &          -0.0001460 \\
  & 1000.0000000       &       0.0053290    &         -0.0011540 \\
  & 228.0000000         &     0.0270770         &    -0.0057250 \\
  & 64.7100000          &    0.1017180         &    -0.0233120 \\
  &  21.0600000          &    0.2747400        &     -0.0639550 \\
  &   7.4950000          &    0.4485640        &     -0.1499810 \\
  &  2.7970000         &     0.2850740      &       -0.1272620 \\
  &  0.5215000          &    0.0152040   &           0.5445290 \\
C&    S& & \\
      & 0.1596000 &              1.0000000 & \\
C &   P& & \\
      & 9.4390000  &            0.0381090 & \\
      & 2.0020000 &             0.2094800 & \\
      & 0.5456000 &             0.5085570 & \\
      \hline
C &   P& & \\
     & 0.1467000  &             1.0000000 &\\
      \hline
C  &  D &  & \\
     &  0.5500000  &           1.0000000 & \\
\end{tabular}
\caption{\label{bas:custom-cc-pVDZ-basis} Custom cc-pVDZ basis for C atoms in NWChem format. This basis is used for all \mbox{H-Ti-C$_n$-H} chain systems unless otherwise stated.
 %The main exception is for the most compressed bond length of \mbox{H-Ti-C$_{29}$-H} and for \mbox{H-Ti-C$_7$-H} at the same bondlength (only in Fig.~\ref{fig:H-Ti-Cn-H-PEC}).
  All basis set functions are standard except for the P function between the horizontal lines. The Gaussian exponent has been decreased from the original value of 0.151700. }
\end{figure}

\begin{figure}[htb!]
\begin{tabular}{l l l l }
C&    S& & \\
  & 6665.0000000       &       0.0006920   &          -0.0001460 \\
  & 1000.0000000       &       0.0053290    &         -0.0011540 \\
  & 228.0000000         &     0.0270770         &    -0.0057250 \\
  & 64.7100000          &    0.1017180         &    -0.0233120 \\
  &  21.0600000          &    0.2747400        &     -0.0639550 \\
  &   7.4950000          &    0.4485640        &     -0.1499810 \\
  &  2.7970000         &     0.2850740      &       -0.1272620 \\
  &  0.5215000          &    0.0152040   &           0.5445290 \\
C&    S& & \\
      & 0.1596000 &              1.0000000 & \\
C &   P& & \\
      & 9.4390000  &            0.0381090 & \\
      & 2.0020000 &             0.2094800 & \\
      & 0.5456000 &             0.5085570 & \\
      \hline
      \hline
C &   P& & \\
     & 0.1457000  &             1.0000000 &\\
      \hline
      \hline
C  &  D &  & \\
     &  0.5500000  &           1.0000000 & \\
\end{tabular}
\caption{\label{bas:custom-cc-pVDZ-basis-1.6} Custom cc-pVDZ basis for C atoms in NWChem format. Similar to Figure~\ref{bas:custom-cc-pVDZ-basis} with the gaussian exponenet further modifed in order to remove near linear dependence at b=1.6 \r{A} in \mbox{H-Ti-C$_{29}$-H}.
This basis is also used for \mbox{H-Ti-C$_{7}$-H} at b=1.6 \r{A} in Fig.~\ref{fig:H-Ti-Cn-H-PEC} for comparison purposes.
}
\end{figure}

\begin{figure}[!]
\begin{tabular}{l l l l }
C &   S & & \\
 &  8236.0000000            &  0.0005310           &  -0.0001130 \\
  & 1235.0000000          &    0.0041080          &   -0.0008780 \\
    &280.8000000         &     0.0210870         &    -0.0045400 \\
     &79.2700000        &      0.0818530        &     -0.0181330 \\ 
     &25.5900000      &        0.2348170      &       -0.0557600 \\
     & 8.9970000     &         0.4344010     &        -0.1268950 \\
      &3.3190000    &          0.3461290   &          -0.1703520 \\
      &0.3643000   &          -0.0089830 &             0.5986840 \\
C &   S & & \\
     & 0.9059000    &          1.0000000 & \\
C  &  S  & & \\
    &  0.1285000     &         1.0000000 & \\
C  &  P & & \\
     &18.7100000     &         0.0140310 & \\
     & 4.1330000     &         0.0868660 & \\
      &1.2000000    &          0.2902160 & \\
C  &  P & & \\
    &  0.3827000   &           1.0000000 & \\
\hline
C &   P &  & \\
    & 0.1609000  &            1.0000000 & \\
\hline
C  &  D & & \\
      &1.0970000 &              1.0000000 & \\
C  &  D & & \\
     & 0.3180000    &          1.0000000 & \\
C  &  F & & \\
    &  0.7610000   &           1.0000000 & \\
\end{tabular}
\caption{\label{bas:custom-cc-pVTZ-basis} Custom cc-pVTZ basis for C atoms in NWChem format. All basis set functions are standard except for the P function between the horizontal lines. The Gaussian exponent has been increased from the original value of 0.120900. }
\end{figure}

\begin{figure}[htb!]
\begin{tabular}{l l l l }
C2&    S& & \\
  & 6665.0000000       &       0.0006920   &          -0.0001460 \\
  & 1000.0000000       &       0.0053290    &         -0.0011540 \\
  & 228.0000000         &     0.0270770         &    -0.0057250 \\
  & 64.7100000          &    0.1017180         &    -0.0233120 \\
  &  21.0600000          &    0.2747400        &     -0.0639550 \\
  &   7.4950000          &    0.4485640        &     -0.1499810 \\
  &  2.7970000         &     0.2850740      &       -0.1272620 \\
  &  0.5215000          &    0.0152040   &           0.5445290 \\
C2&    S& & \\
      & 0.1596000 &              1.0000000 & \\
C2 &   P& & \\
      & 9.4390000  &            0.0381090 & \\
      & 2.0020000 &             0.2094800 & \\
      & 0.5456000 &             0.5085570 & \\
C2 &   P& & \\
      & 0.1467000  &             1.0000000 & \\
\end{tabular}
\caption{\label{bas:custom-cc-pVDZ-basis-substrate} Custom basis for C atoms in the \mbox{H$_{22}$C$_{56}$} planar graphitic substrate system in NWChem format. We use the atomic symbol ``C2'' for Carbon atoms which are treated with this basis. The basis set is based on the custom cc-pVDZ basis, given in \ref{bas:custom-cc-pVDZ-basis}, used for \mbox{H-Ti-C$_{29}$-H}  and \mbox{H-Ti-C$_{7}$-H} with the additional modification that the d-states are discarded in order to reduce computational cost. 
%The basis for C atoms within \mbox{H$_{22}$C$_{56}$} closely resembles a 6-31g basis.
This basis closely resembles a 6-31g basis.
}
\end{figure}

\begin{figure}
\begin{center}
\includegraphics[width=0.7\textwidth]{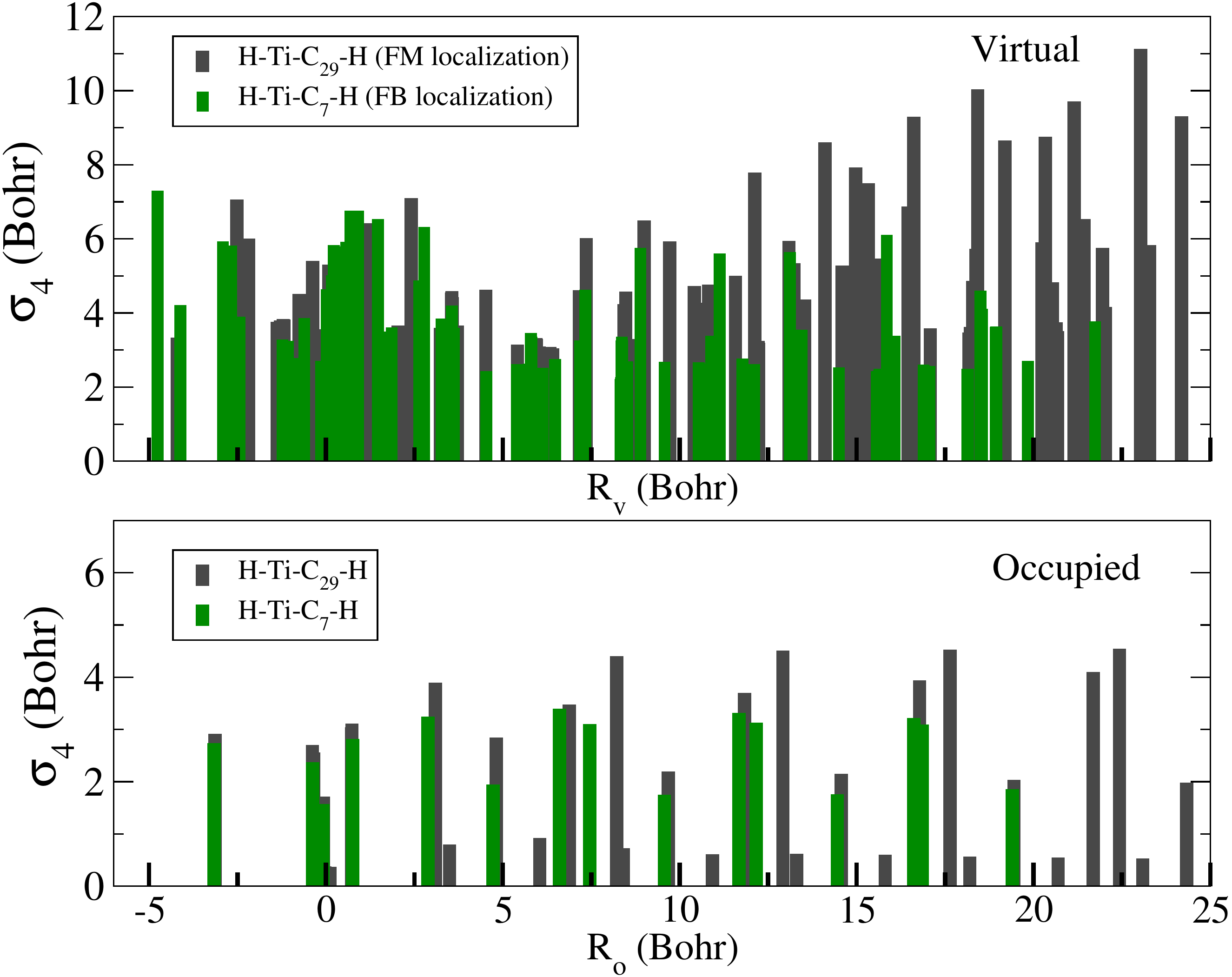}
%\caption{\label{fig:H-Ti-C29-H-centroids4thMom} 4th central moment orbital spread plotted against orbital centroid position along the chain system. The origin is at the equilibrium Ti atom position.}
\caption{
\label{fig:H-Ti-Cn-H-centHist} \mbox{H-Ti-C$_{n}$-H}  
bar plot comparing the fourth central moment $\sigma_4$ [Eq.~(\ref{eq:pthMomSpread})] of the localized virtual (upper panel) and occupied (lower panel) orbitals with $n=7$ and $n=29$
vs. the orbital centroid position along the chain.  
The origin is set to the position of the Ti atom.
The full length of the $n=7$ is shown, while approximately one third of the $n=29$ chain is shown.
}
\end{center}
\end{figure}